\documentclass[onecolumn]{emulateapj}

\usepackage{subfigure,epsfig}
\usepackage{natbib}
\usepackage{rotating}

\newcommand{\Fig}[1]{Fig \ref{#1}}

\newcommand{\beq}{\begin{eqnarray}}
\newcommand{\eeq}{\end{eqnarray}}

\def\lsim{\mathrel{\rlap{\lower4pt\hbox{\hskip1pt$\sim$}} \raise1pt\hbox{$<$}}}                
\def\gsim{\mathrel{\rlap{\lower4pt\hbox{\hskip1pt$\sim$}}\raise1pt\hbox{$>$}}}                
\shorttitle{TTVs: Inclined \& Retrograde Systems}
\shortauthors{Payne,\,Ford\,\& Veras}

\begin{document}

\title{Transit Timing Variations for Inclined and Retrograde Exoplanetary Systems}

%

\author{Matthew J. Payne, Eric B. Ford and Dimitri Veras}
\affil{Department of Astronomy, University of
  Florida, 211 Bryant Space Science Center, P.O. Box 112055,
  Gainesville, FL 32611-2055}
\email{matthewjohnpayne@gmail.com}

%

%
%
\begin{abstract}
We perform numerical calculations of the expected transit timing variations (TTVs) induced on a Hot-Jupiter by an Earth-mass perturber. Motivated by the recent discoveries of retrograde transiting planets, we concentrate on an investigation of the effect of varying relative planetary inclinations, up to and including completely retrograde systems. 
We find that planets in low order (E.g. 2:1) mean-motion resonances (MMRs) retain approximately constant TTV amplitudes for $0<\,^{\circ}i<170\,^{\circ}$, only reducing in amplitude for $i > 170\,^{\circ}$.
Systems in higher order MMRs (E.g. 5:1) increase in TTV amplitude as inclinations increase towards $45\,^{\circ}$, becoming approximately constant for $45 < i < 135$, and then declining for $i > 135\,^{\circ}$.
Planets away from resonance slowly decrease in TTV amplitude as inclinations increase from 0 to 180, where-as planets adjacent to resonances can exhibit a huge range of variability in TTV amplitude as a function of both eccentricity and inclination.
For highly retrograde systems ($135\,^{\circ} < i \leq 180\,^{\circ}$), TTV signals will be undetectable across almost the entirety of parameter space, with the exceptions occurring when the perturber has high eccentricity or is very close to a MMR.
This high inclination decrease in TTV amplitude (on and away from resonance) is important for the analysis of the known retrograde and multi-planet transiting systems, as inclination effects need to be considered if TTVs are to be used to exclude the presence of any putative planetary companions: absence of evidence is not evidence of absence.
\end{abstract}

%
%

\section{Introduction}
Of the known extra-Solar planets, more than 60 transit the host star. Of these systems, at least four show evidence for an external companion (GJ 436, \citealt{Maness07}; HAT-P-13, \citealt[][]{Bakos09}; HAT-P-7, \citealt[][]{Winn09} and CoRoT-7 \citealt[][]{Queloz09}).

If the transiting planet were the only planet in the system, then the period between each successive transit would be constant (neglecting complicating effects such as general relativity, stellar oblateness and tides). The presence of additional planets in the system (which themselves may or may not transit the star) can cause perturbations to the orbit of the transiting planet, leading to detectable transit timing variations (TTVs) of the known transiting planet \citep{Miralda-Escude02,Holman05,Agol05}. Other studies have extended the TTV method to demonstrate the feasibility of detecting planetary moons \citep{Kipping09a,Kipping09b} and Trojan companions \citep{Ford07b} in extra-solar planetary systems.

However, practical use of TTVs as a detection tool requires the solution of the difficult ``inverse problem'', i.e. given a particular TTV profile, can one reconstruct (or at least restrict) the mass and orbit of the unseen perturber? This problem is non-trivial, as numerous different perturber mass-orbit configurations can lead to degenerate TTV solutions \citep[e.g.][]{Nesvorny08}. 

Several investigations have considered (and ruled out) planets more massive than the Earth in certain orbits / regions of parameter space close to many of the known transiting planets \citep[e.g. ][]{Steffen05,Agol07,Alonso08,Bean08}, while \citet{Nesvorny08} and \citet{Nesvorny09} have developed an approximate analytic method to try and tackle the inverse problem in a more general manner.

However, most previous investigations have focused on analyzing the effect of prograde, coplanar companions. Even where explicit investigations of inclination effects have been conducted \citep[e.g.][]{Nesvorny09}, the investigations have been restricted to prograde cases, examining only relative inclinations significantly lower than $90\,^{\circ}$.

There are several reasons to consider TTVs of inclined systems. Observationally, it is possible to measure the sky-projected angle between the spin vector of a star, and the orbital angular momentum vector of a planet transiting that star via the Rossiter-McLaughlin effect. Measurements over the past year have revealed that a number of the known transiting planets are strongly inclined, or even retrograde (e.g. HAT-P-7b, \citealt{Narita09} and \citealt{Winn09},  as well as WASP-17b, \citealt{Anderson09}).

Although the mechanism(s) driving the creation of such retrograde planets remains unclear, models used to explain high \emph{eccentricity} exo-planets via planet-planet scattering \citep{Rasio96b,Weidenschilling96,Lin97b,Levison98,Papaloizou01,Moorhead05,Chatterjee08,Juric08,Ford08a} and/or Kozai oscillations \citep[E.g.][]{Kozai62,Takeda05,Nagasawa08} can naturally excite large orbital \emph{inclinations} (although the quantitative details of the distributions can differ greatly). Other mechanisms such as inclination-pumping during migration \citep[e.g.][]{Lee09} may also contribute.

In addition, previous dynamical studies have suggested that systems of prograde and retrograde planets are \emph{more} stable than standard prograde-prograde cases \citep{Gayon08,Gayon09,Smith09}, essentially because the planets spend less time ``close together'' and thus perturbations are smaller.  

Given the observational evidence for highly inclined and retrograde systems, and the suggestion of enhanced stability in such systems, we investigate  and quantify the hypothesis that retrograde systems will have a significantly reduced TTV profile compared to a standard prograde case.


\section{Methodology}
\subsection{Integration and TTVs}
We consider a fiducial system consisting of two planets: a transiting Hot-Jupiter (with mass $M_1=M_{Jupiter}$, eccentricity $e_1=0$ and semi-major axis $a_1=0.05\,AU$) and a small non-transiting, Earth-mass planet which perturbs the transit times of the larger planet. We evolve each system forward in time for $\sim 10$ years, equating to $\sim 1,000$ transits of the Jupiter-mass planet (assuming that Hot-Jupiter remains transiting throughout).

We perform the n-body integrations using a conservative Bulirsch-Stoer integrator, derived from that of {\sc Mercury} (Chambers 1999).  We use a barycentric coordinate system and limit the time steps to no more than 0.04 times the orbital period.  After each time step, we test whether the projected star-planet separation ($\Delta$) passed through a local minimum and the planet in question is closer to the observer than the star.  Each time these conditions are met, we find the nearby time that minimizes $\Delta$ via Newton-Raphson iteration and increment an index $i$.  If the minimum $\Delta$ is less than the stellar radius, then we record the mid-time of the transit, $t_{i}$.  Observations of transit times are perturbed by the light travel time, $\delta~t_{ltt}(i) \simeq -({\bf r}_p(t_{i}) \cdot {\bf r}_{los})/c$, where ${\bf r}_p(t_i)$ is the barycentric vector of the planet at time $t_i$, ${\bf r}_{los}$ is a unit vector pointing to the observer, and $c$ is the speed of light.  The observable transit time variations are calculated as $\delta~t(i) = t_{i}+\delta~t_{llt}(i)- i\times\hat{P}-\hat{t}_0$, where the constant $\hat{P}$ and $\hat{t}_0$ are determined by linear least squares minimization of $\sum_i (\delta~t(i))^2$. We neglect any motion of the stellar center between the time of light emission and the time of transit.

\subsection{TTV Maps}
In \S \ref{ResultsFixedMaps} and \ref{ResultsVarying} we produce TTV contour maps in $(i_2,e_2)$ and $(\frac{a_2}{a_1},e_2)$ parameter-space. These plots are produced by fixing the mass and the initial orbit of the transiting hot-Jupiter ($e_1=0$ \& $a_1=0.05$), and then varying the orbit of the Earth-mass perturber. 

For each set of configurations - i.e. semi-major axis, $a_2$, eccentricity, $e_2$, \& inclination, $i_2$ of the perturbing planet - we conduct 5 simulations randomising the other orbital elements (argument of pericentre, $\omega$, longitude of ascending node, $\Omega$, \& initial mean anomaly $M$). 

We then calculate the RMS of $\delta t(i)$ for each $a,e,i$ configuration. The required contour plots are then plotted using an approximately logarithmic color scale (shown in Figs. \ref{FIG:Fiducial_Inc_Scan} and \ref{FIG:Fiducial_Inc_1}).

We restrict our simulations to the region defined by the prograde coplanar three-body Hill stability criterion of \citet{Gladman93}. This ensures that all of the $i=0\,^{\circ}$ simulations are stable, and ensures for the $i>0\,^{\circ}$ systems that we have covered all of the guaranteed stability zone \emph{and more} (see \citet{Veras04} for a discussion of the stability limit as a function of inclination). As an additional check on systems with extremely large TTVs close to the Hill-stability criterion boundary, we conducted 1,000 longer-term {\sc Mercury} simulations. In most cases the Earth-mass planet experiences significant changes in semi-major axis within 1,000 years, due to the ``kicks'' at pericentre passage. Thus, although these planets do not collide, the less massive planet can experience some orbit variation.


\section{Results}\label{Results}
\subsection{Example TTVs For a Hot-Jupiter System}\label{ResultsFixedExample}
%
\begin{figure*}
  \centering
  \begin{tabular}{ccc}
    \includegraphics[angle=-90,width=0.31\textwidth]{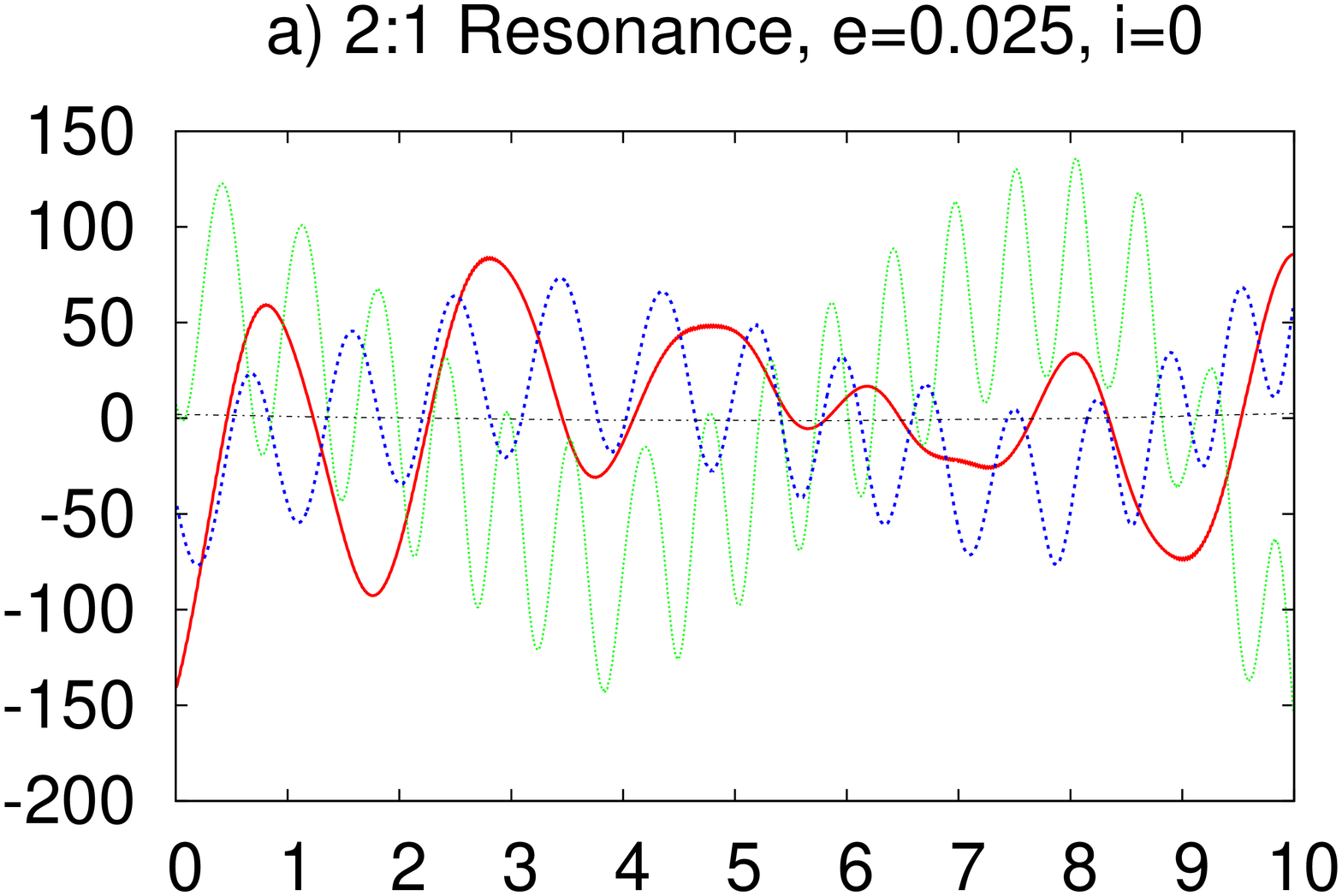}&
    \includegraphics[angle=-90,width=0.31\textwidth]{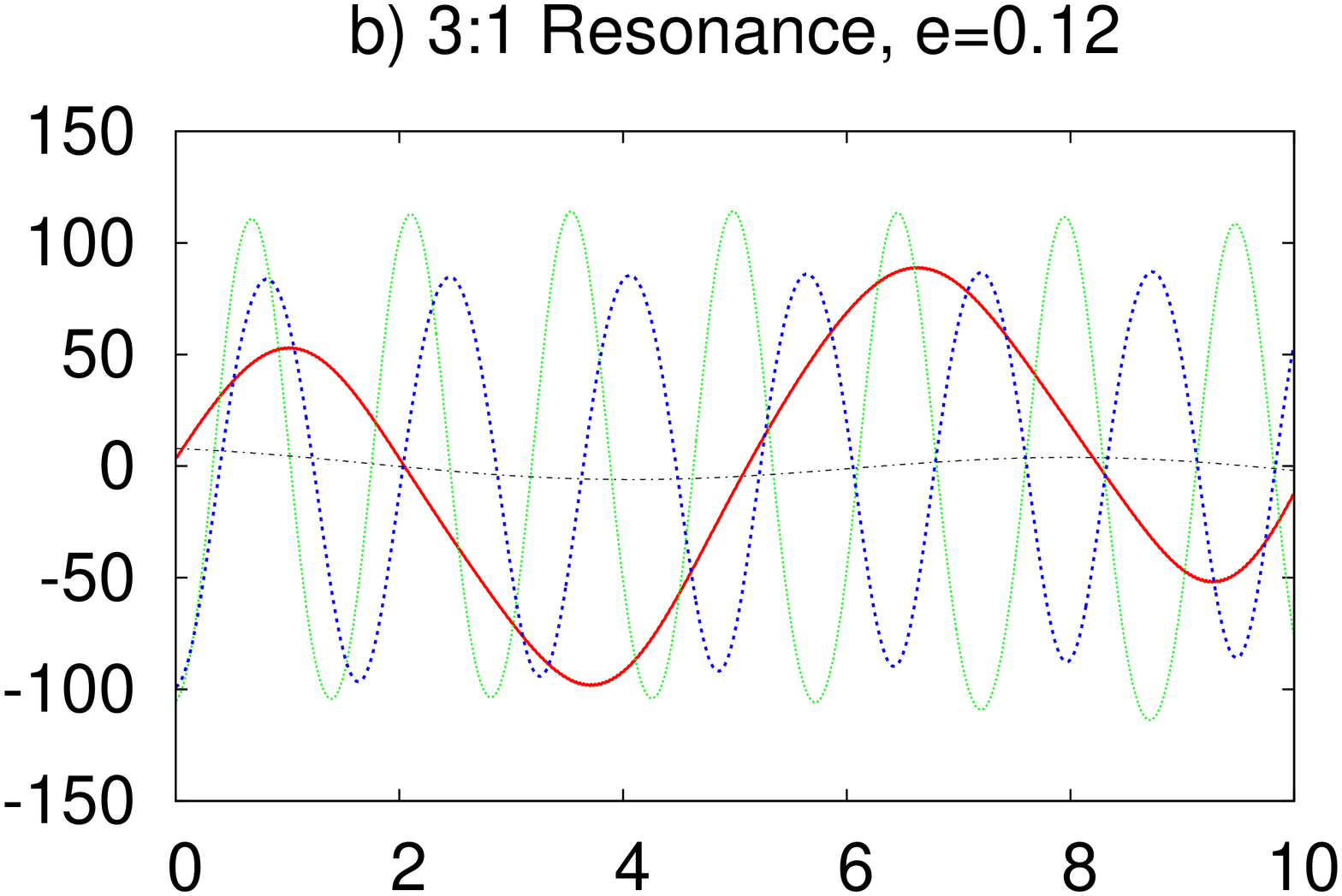}&
    \includegraphics[angle=-90,width=0.31\textwidth]{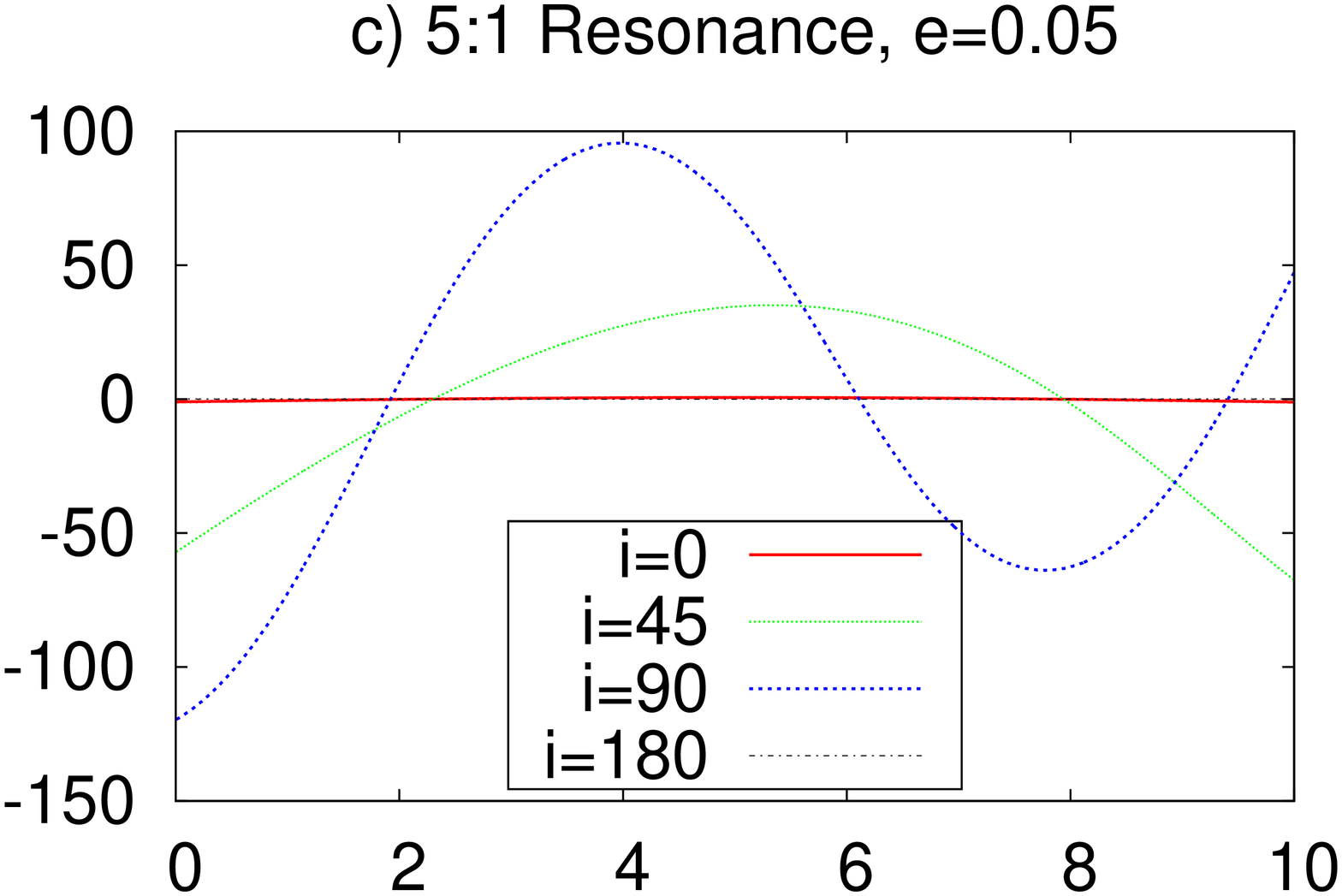}\\

    \includegraphics[angle=-90,width=0.31\textwidth]{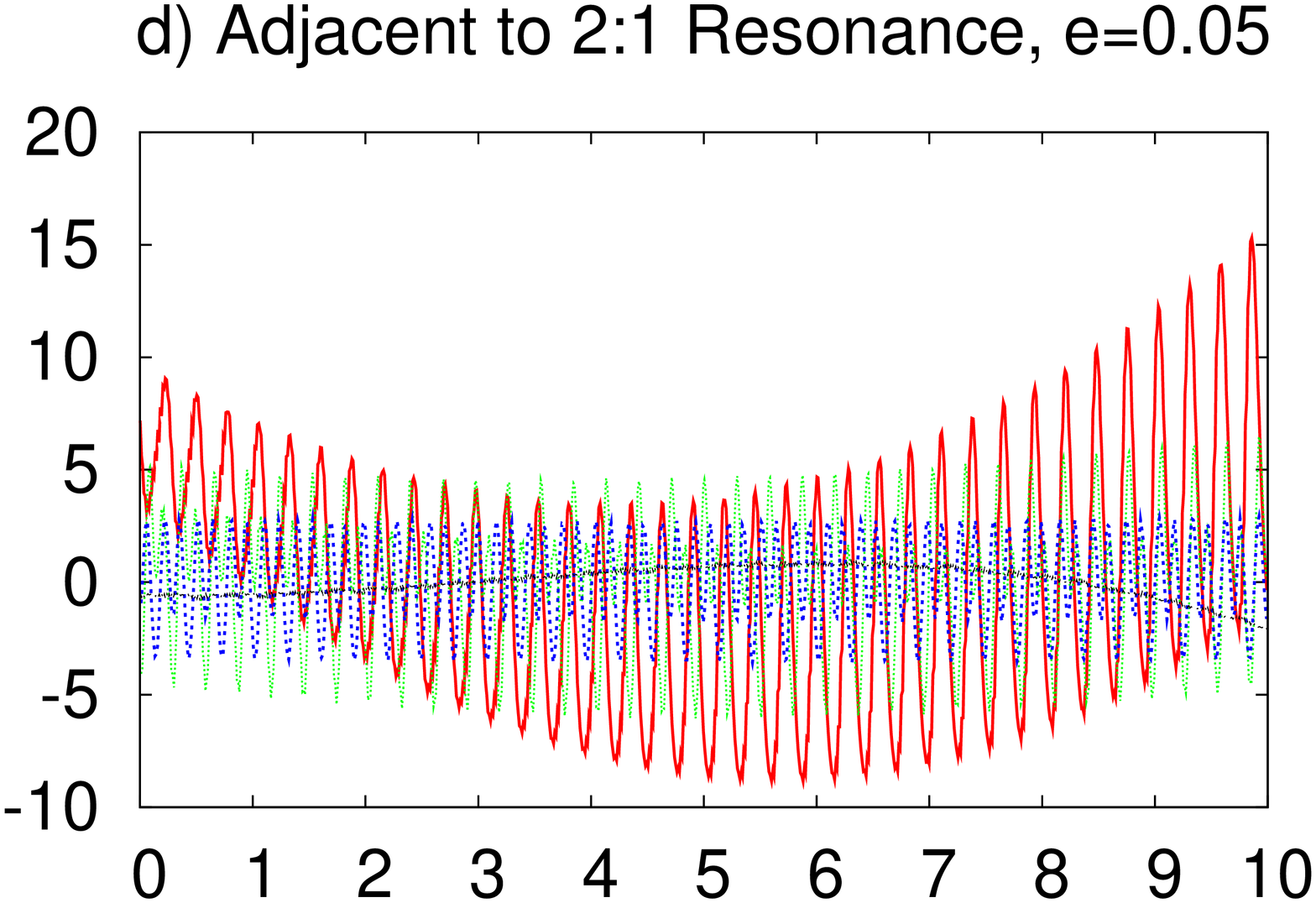}&
    \includegraphics[angle=-90,width=0.31\textwidth]{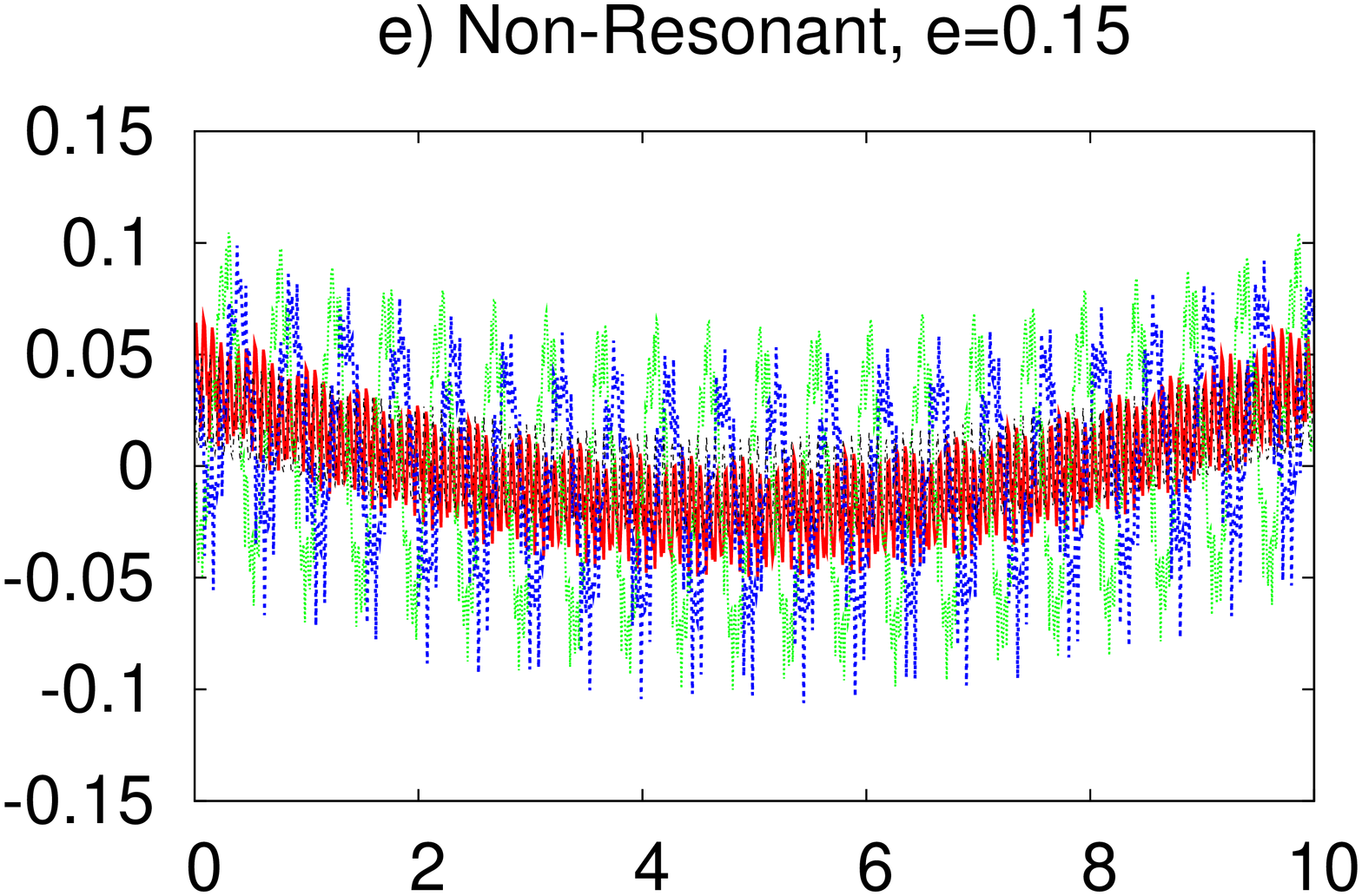}&
    \includegraphics[angle=-90,width=0.31\textwidth]{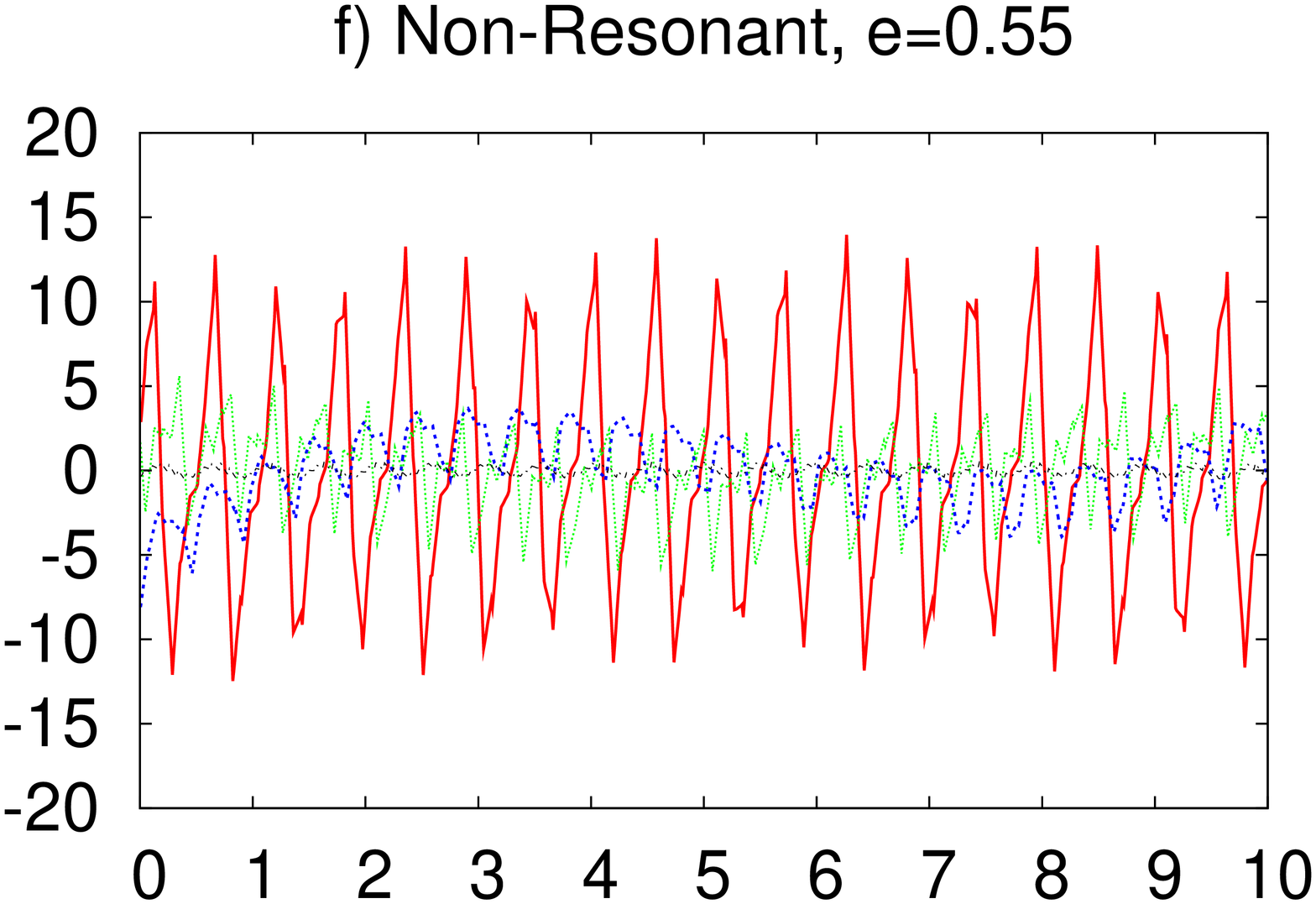}\\
  \end{tabular}
  \put(-515,-70){{\begin{sideways} \centering {\bf RMS TTV (seconds)} \end{sideways}}}
  \put(-280,-130){\centering \bf Time (years)}
  \caption{The effect of varying inclination at fixed semi-major axes and eccentricities. Prograde orbits with: $i=0\,^{\circ}$ have solid red lines; $i=45\,^{\circ}$ have dotted green lines; $90\,^{\circ}$ have dashed blue lines; while retrograde $i=180\,^{\circ}$ orbits have dot-dashed black lines. In (a) we show the TTVs for a system on/near the 2:1 MMR ($a_2 / a_1 = 1.59$,\,$P_2 / P_1 = 2.0$,\,$e_2 = 0.02$), In (b) the system is on/near the 3:1 MMR ($a_2 / a_1 = 2.08$,\,$P_2 / P_1 = 3.0$,\,$e_2 = 0.12$), In (c) the system is on/near the 5:1 MMR ($a_2 / a_1 = 2.92$,\,$P_2 / P_1 = 5.0$,\,$e_2 = 0.05$), In (d) we the system is just outside the 2:1 MMR ($a_2 / a_1 = 1.63$,\,$P_2 / P_1 = 2.08$,\,$e_2 = 0.05$), In (e) we show a  a non-resonant system at low eccentricity ($a_2 / a_1 = 3.6$,\,$P_2 / P_1 = 6.83$,\,$e_2 = 0.15$), In (f) we show a non-resonant system at high eccentricity ($a_2 / a_1 = 3.6$,\,$P_2 / P_1 = 6.83$,\,$e_2 = 0.55$). The results are complex, with different systems exhibiting a number of differing frequency contributions, with many contributions having different dependencies on inclination. However it is clear from these plots that in all the examples shown, the TTV signal for the $i=180\,^{\circ}$ case tends to be significantly lower in amplitude than for other inclinations.}
  \label{FIG:INDIVIDUAL}
\end{figure*}

In \Fig{FIG:INDIVIDUAL}a we show a circular hot-Jupiter at 0.05 au being perturbed by an Earth-mass companion with $e_2 = 0.02$ located on the external 2:1 mean-motion resonance (MMR). We find that although the prograde systems ($0\,^{\circ} \leq i \leq 90\,^{\circ}$) have TTV profiles that differ significantly, their overall amplitudes are similar. In contrast, the amplitude of the $i=180\,^{\circ}$ signal is very low ($\sim 1$ second). The results for the 3:1 case in \Fig{FIG:INDIVIDUAL}b are qualitatively similar.

In contrast, for the 5:1 MMR system (\Fig{FIG:INDIVIDUAL}c), while the amplitude of the $i=180\,^{\circ}$ signal is extremely low, there are now significant differences in the behaviour at smaller angles: the $i=0\,^{\circ}$ and $i=180\,^{\circ}$ now have similarly low amplitudes, with an approximate maximum occurring at $i=90\,^{\circ}$. 

Away from resonance (Figs \ref{FIG:INDIVIDUAL}d, \ref{FIG:INDIVIDUAL}e \& \ref{FIG:INDIVIDUAL}f), we again find that the $i=180\,^{\circ}$ signal is generally significantly lower than that observed for cases with $i<90\,^{\circ}$, but that the detailed dependence on inclination is far from obvious, with some cases (Figs \ref{FIG:INDIVIDUAL}d \& \ref{FIG:INDIVIDUAL}f) having the highest amplitude at $i=0\,^{\circ}$, while the low eccentricity plot of Fig. \ref{FIG:INDIVIDUAL}e has a particularly low amplitude signal at $i=0\,^{\circ}$.

We note that there are a number of frequencies evident in \Fig{FIG:INDIVIDUAL}, with periods from a few months to $>>10$ years. In general, the short-term oscillations tend to be driven by variations in $e_2$, leading to variations in the distance of closest approach between the two planets. The longer term quadratic trends are primarily due to outward drifts in $a_2/a_1$ as a result of kicks at close approach.

The large number of different frequencies evident in the plots also serves to highlight the importance of considering the sampling period over which observations and simulations are to be conducted and compared.

Furthermore, TTV profiles may change in amplitude and frequency by several orders of magnitude simply due to a variation in the initial values of the mean anomalies \citep{Veras09c}. This means that while approximate expressions for TTV amplitude dependencies on mass, semi-major axis separation, etc, can serve as a useful guide, individual system variations can be far removed from the mean. We do not intend this statement as a criticism of previous work, but rather as a cautionary note against over-reliance on approximations when analyzing individual systems which can exhibit great sensitivity to initial conditions. We do not seek in this paper to address in detail all such issues, but instead defer a more detailed consideration to a companion paper \emph{Veras et al. 2010, in prep.}, where a more thorough examination of the coplanar prograde case will be presented, allowing a much more detailed exposition of the huge number of variables which can affect the TTV signal.

\subsection{Eccentricity-Inclination Contour Plots}\label{ResultsFixedMaps}
%
\begin{figure*}
  \centering
  \begin{tabular}{ccc}
    \includegraphics[angle=0,width=0.33\textwidth]{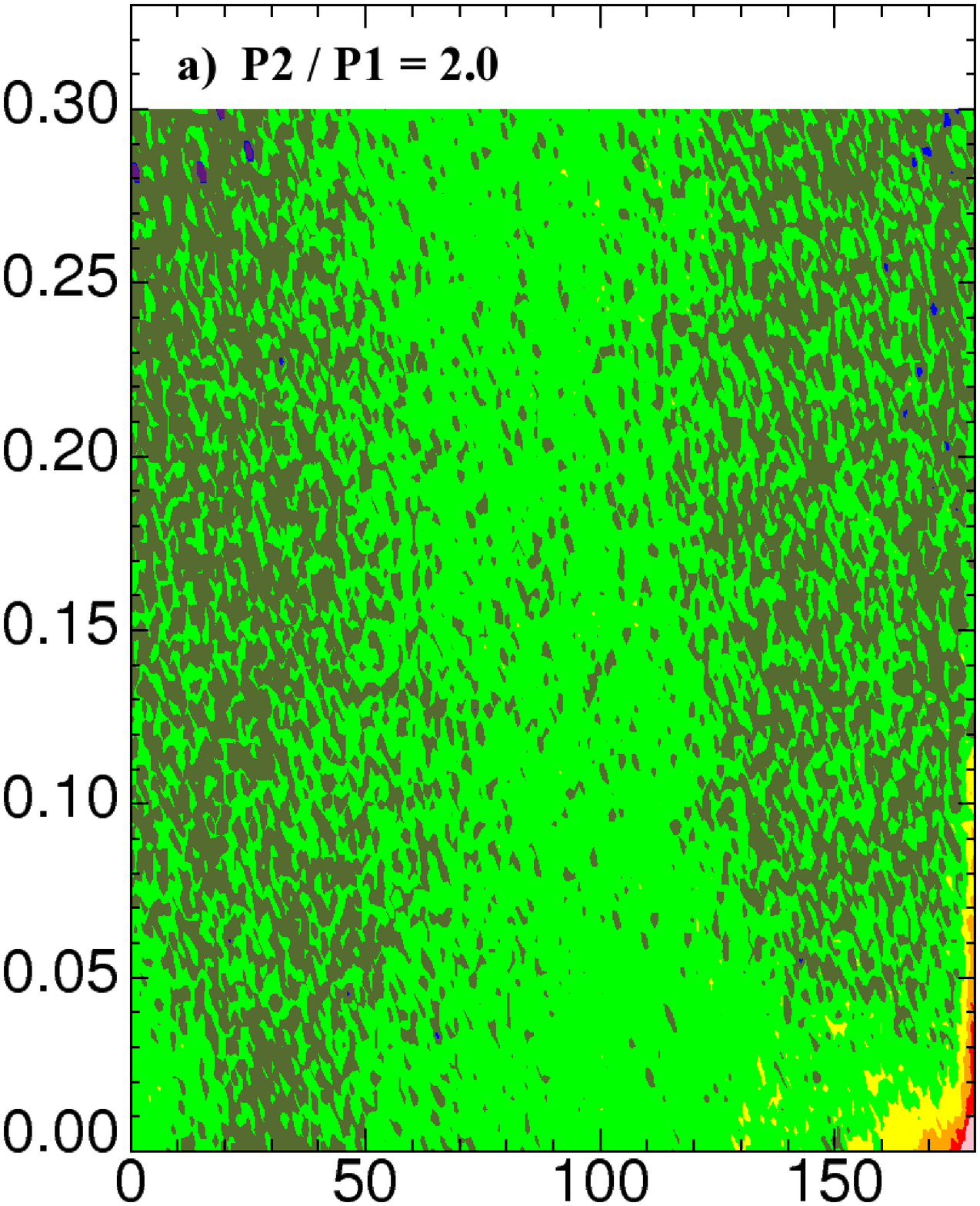}&
    \includegraphics[angle=0,width=0.33\textwidth]{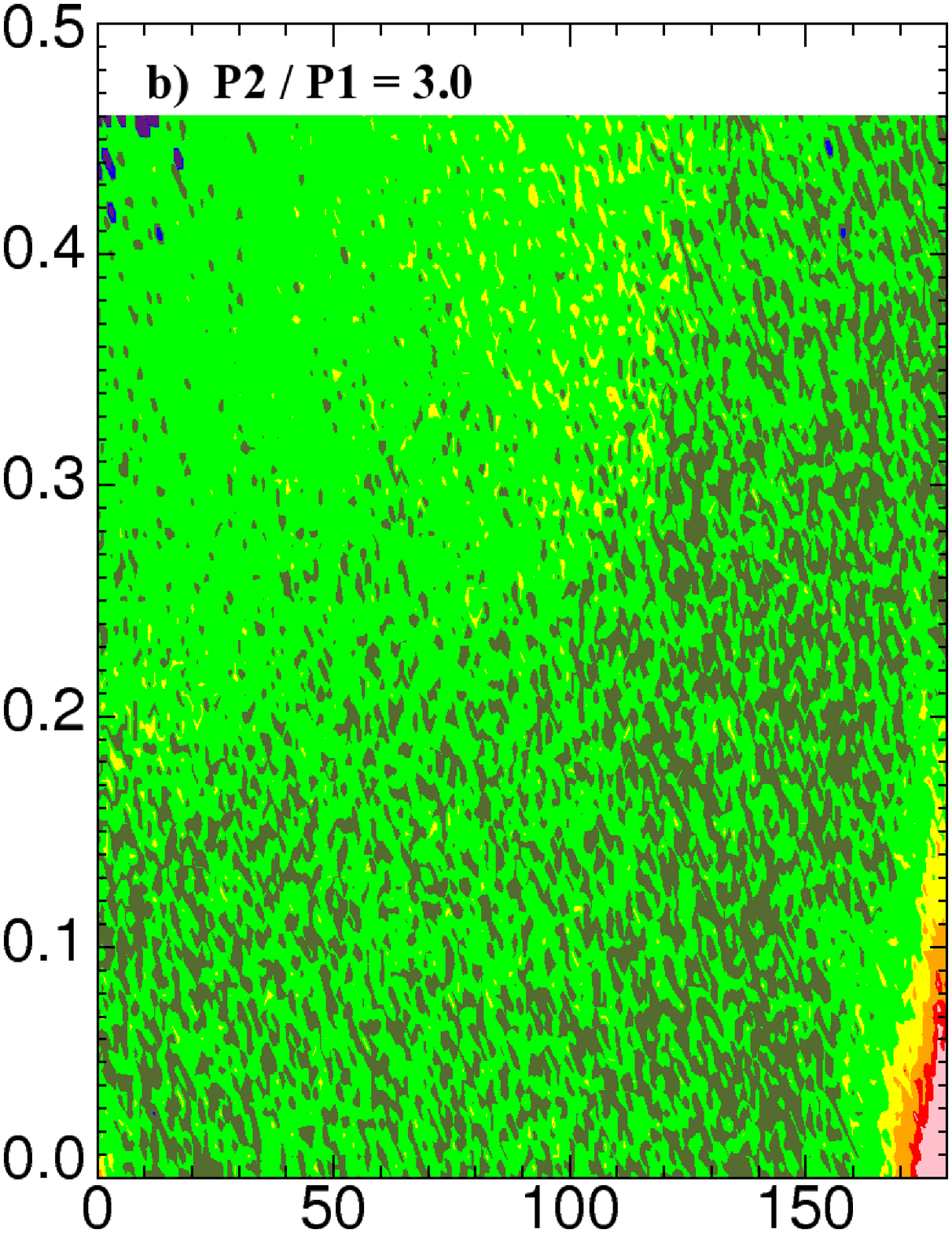}&
    \includegraphics[angle=0,width=0.33\textwidth]{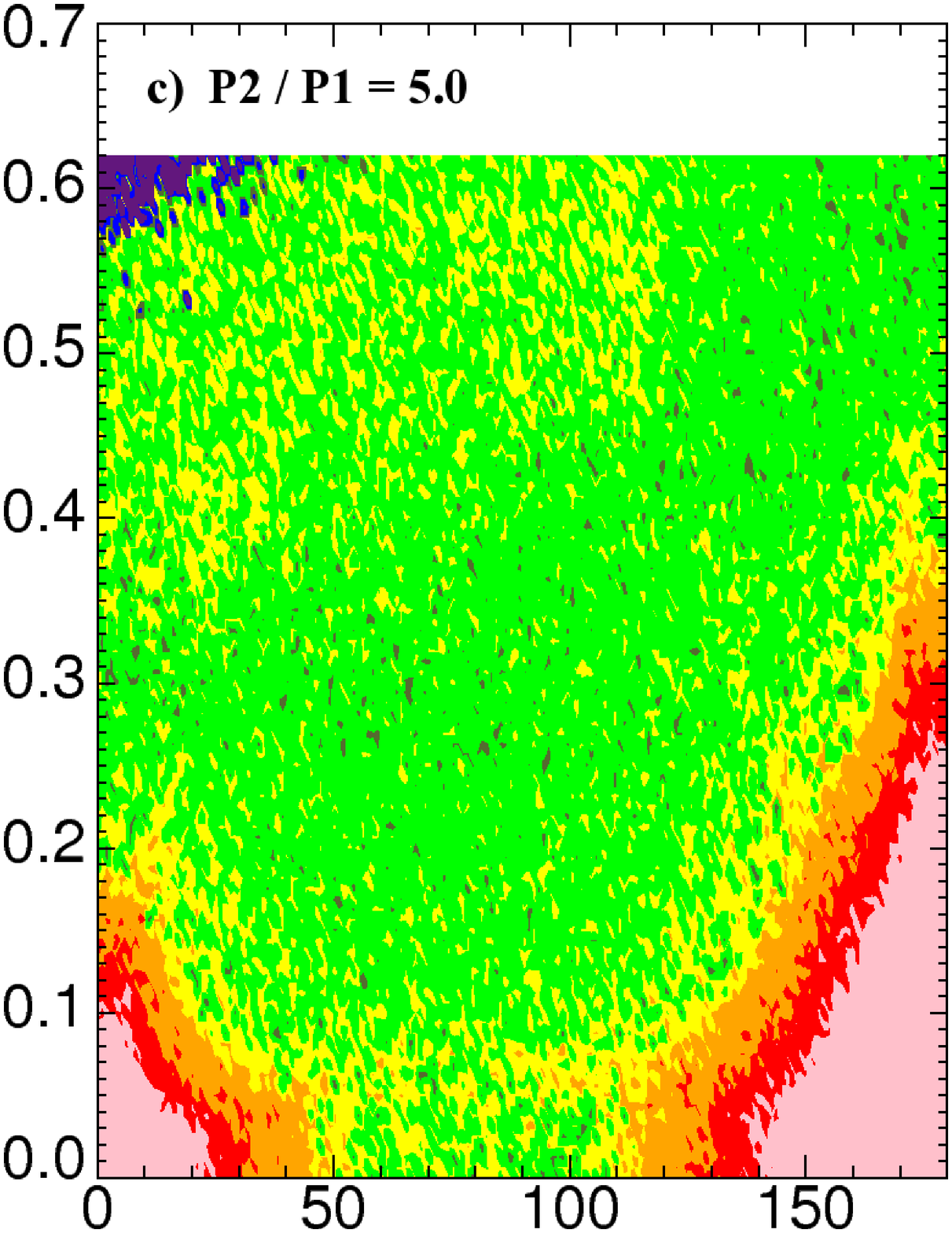}\\
    \includegraphics[angle=0,width=0.33\textwidth]{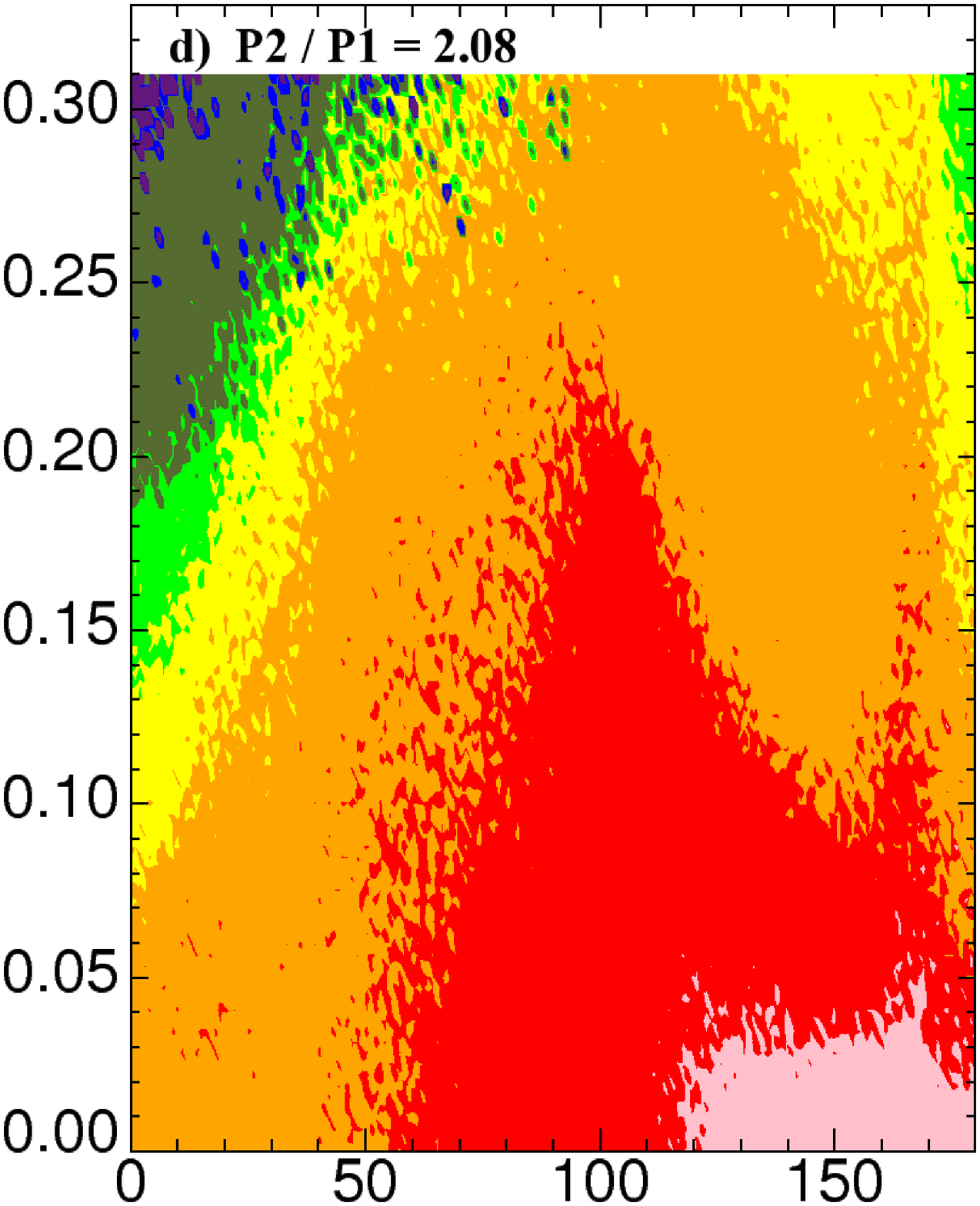}&
    \includegraphics[angle=0,width=0.33\textwidth]{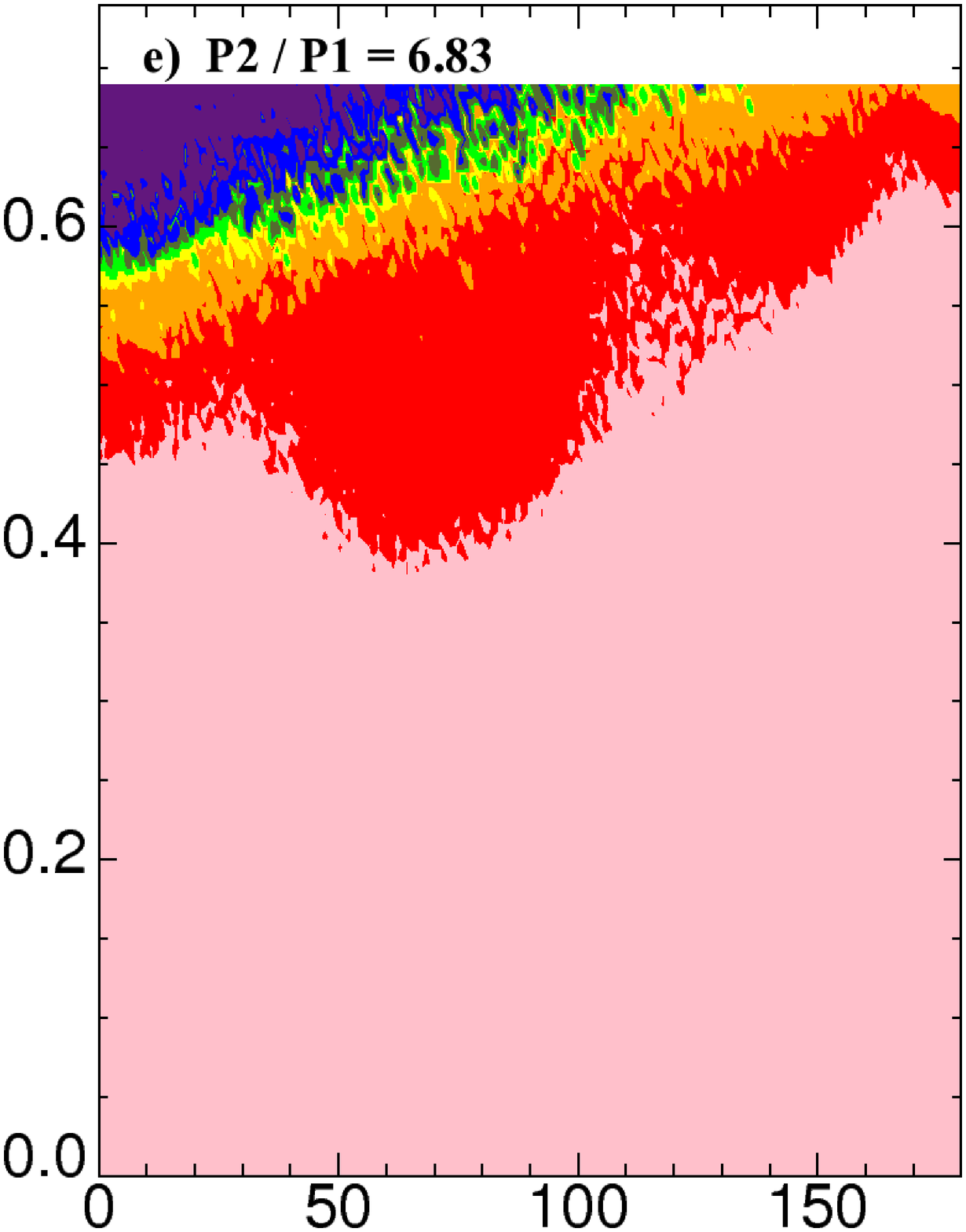}&
    \includegraphics[angle=0,width=0.33\textwidth]{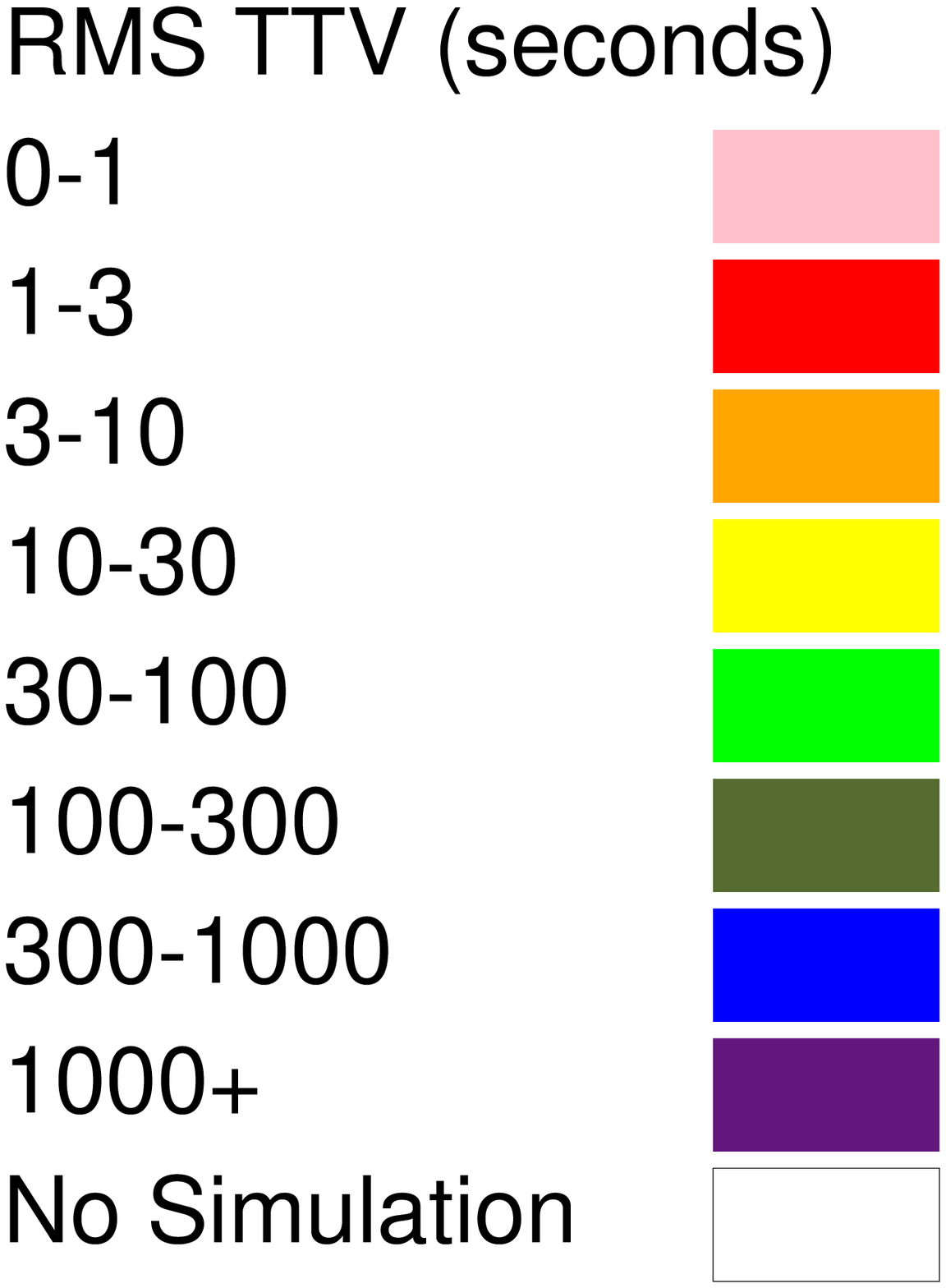}
\\
  \end{tabular}
%
  \put(-545,-120){{\begin{sideways}\parbox{10cm}{\centering \bf Eccentricity of the Outer Body ($e_2$)}\end{sideways}}}
  \put(-350,-230){\centering \bf Relative Inclination of the Outer Body ($i_2$).}
  \caption{Plots at fixed ($a_2/a_1$) showing the median RMS TTV amplitude as a function of $e_2$ \& $i_2$. The contour key is displayed in the bottom right-hand corner. The semi-major axes selected for plots (a) - (e) are the same as those of \Fig{FIG:INDIVIDUAL}a - \Fig{FIG:INDIVIDUAL}e. We can see significant differences in behaviour as a function of inclination depending on (i) position on / off resonance, (ii) the type of any resonance, and (iii) the eccentricity of the the perturber.}
  \label{FIG:Fiducial_Inc_Scan}
\end{figure*}

To better understand the TTV amplitudes as a function of inclination, for each of the semi-major axis plots in \Fig{FIG:INDIVIDUAL} we perform a search of the $(e_2,i_2)$ parameter space, producing contour plots in \Fig{FIG:Fiducial_Inc_Scan} showing the RMS TTV amplitude variation at that semi-major axis separation as the eccentricity and inclination of the outer planet is varied.

In the 2:1 case of \Fig{FIG:Fiducial_Inc_Scan}a the resulting contour plot is rather uniform, with the only area of significant reduction being in the $e_2 < 0.1$, $i\sim 180\,^{\circ}$ region. An approximately similar result can be seen for the 3:1 case.

For the 5:1 case (\Fig{FIG:Fiducial_Inc_Scan}c), the situation changes: the region of low amplitude close to $i = 180\,^{\circ}$ significantly expands, while another low amplitude region emerges at $e_2 < 0.1$ \& $i < 45\,^{\circ}$. 

For \Fig{FIG:Fiducial_Inc_Scan}d the perturber is located very close to (but not on) the 2:1 MMR. The plotted behavior is particularly rich, showing huge variations in the predicted TTV amplitude as functions of both eccentricity and inclination.

Finally, the non-resonant case in \Fig{FIG:Fiducial_Inc_Scan}e with $a_2/a_1 = 3.6$, shows a fairly simple decline in TTV amplitude as a function of inclination.

\subsection{TTV Maps for Varying $\frac{a_2}{a_1}$}\label{ResultsVarying}
We now look at a different projection of the data, looking in the $(\frac{a_2}{a_1},e_2)$ plane at the TTV structure as a function of inclination.


\begin{figure*}
  \centering
  \begin{tabular}{ccc}
    \includegraphics[angle=0,width=0.33\textwidth]{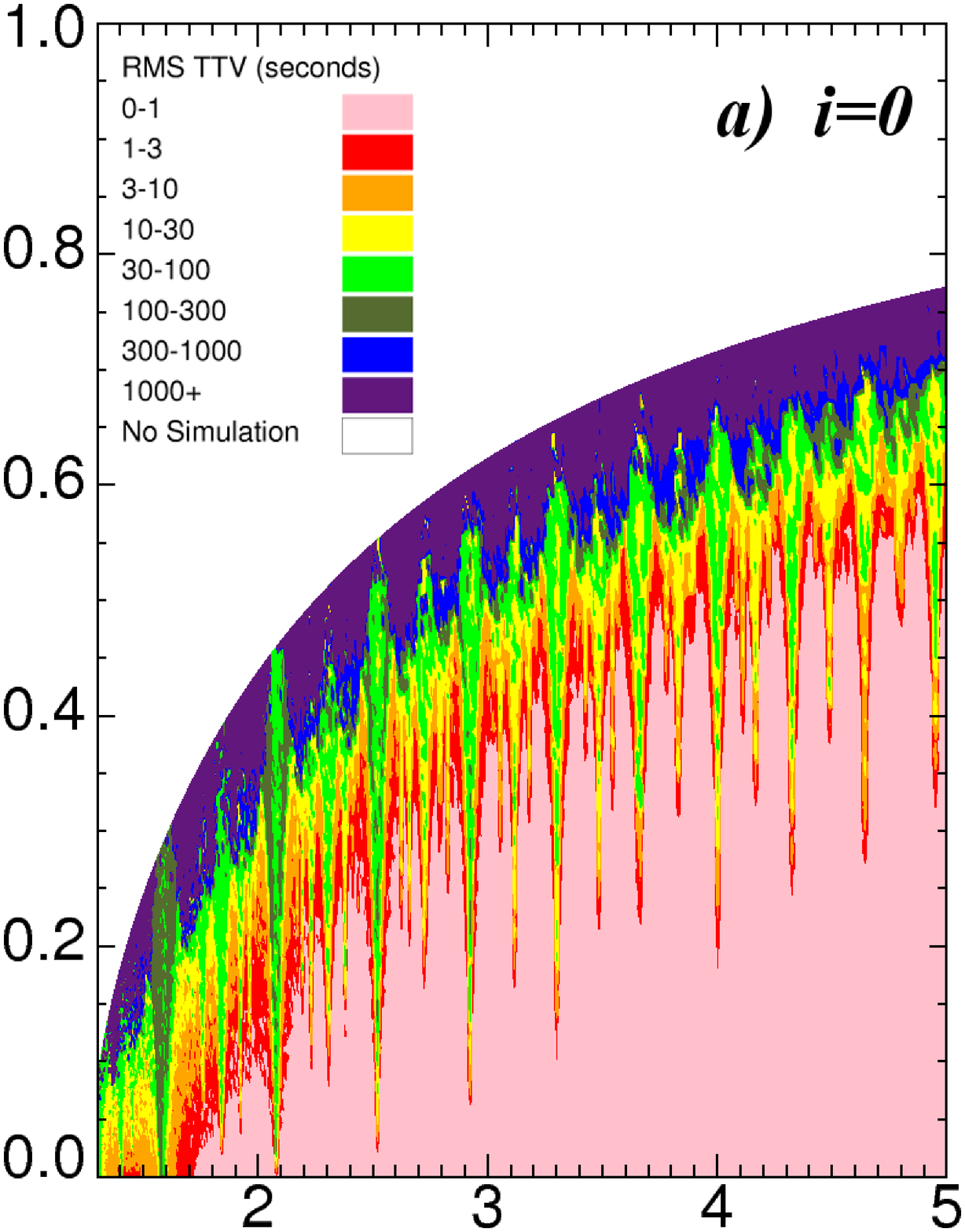}&
    \includegraphics[angle=0,width=0.33\textwidth]{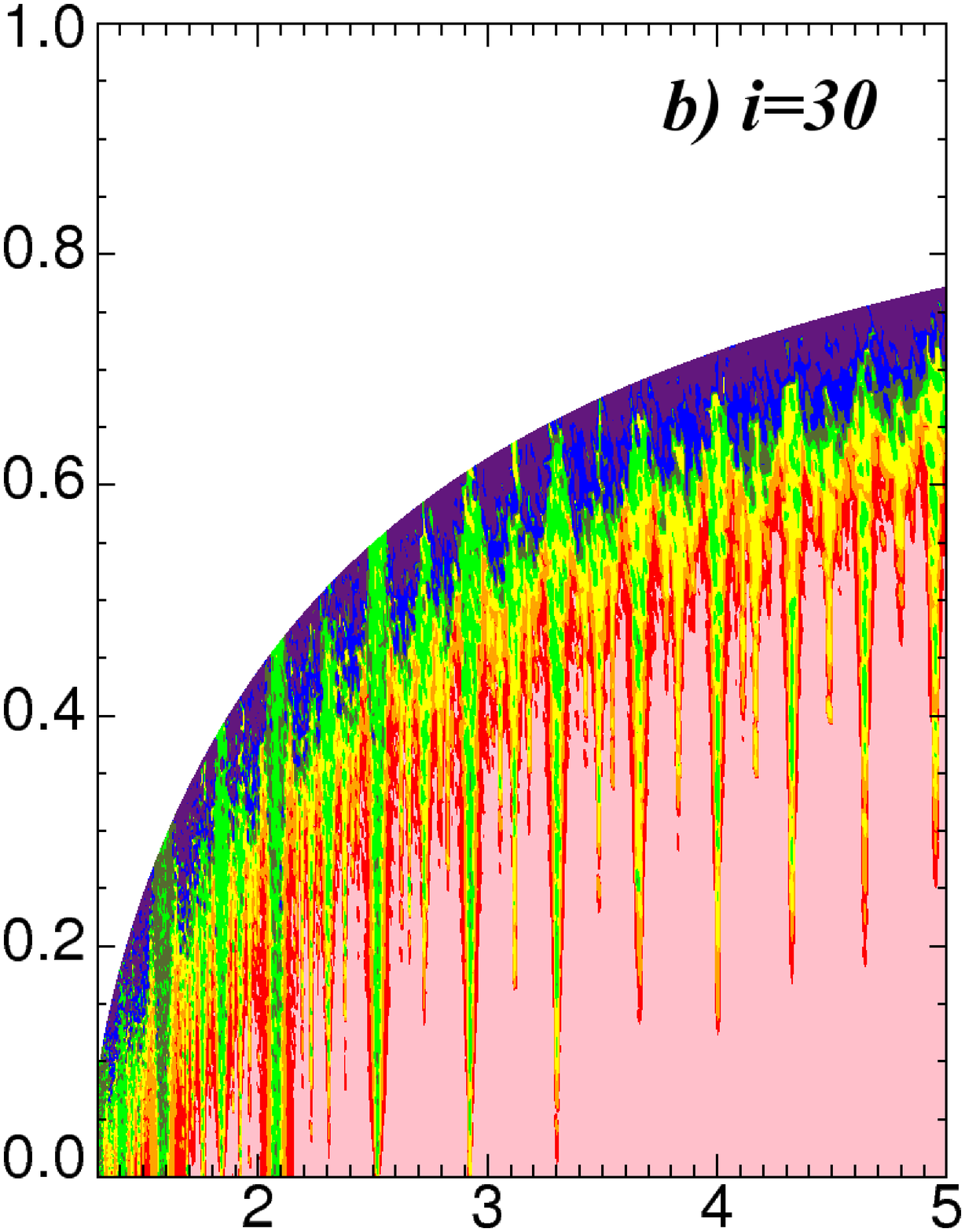}&
    \includegraphics[angle=0,width=0.33\textwidth]{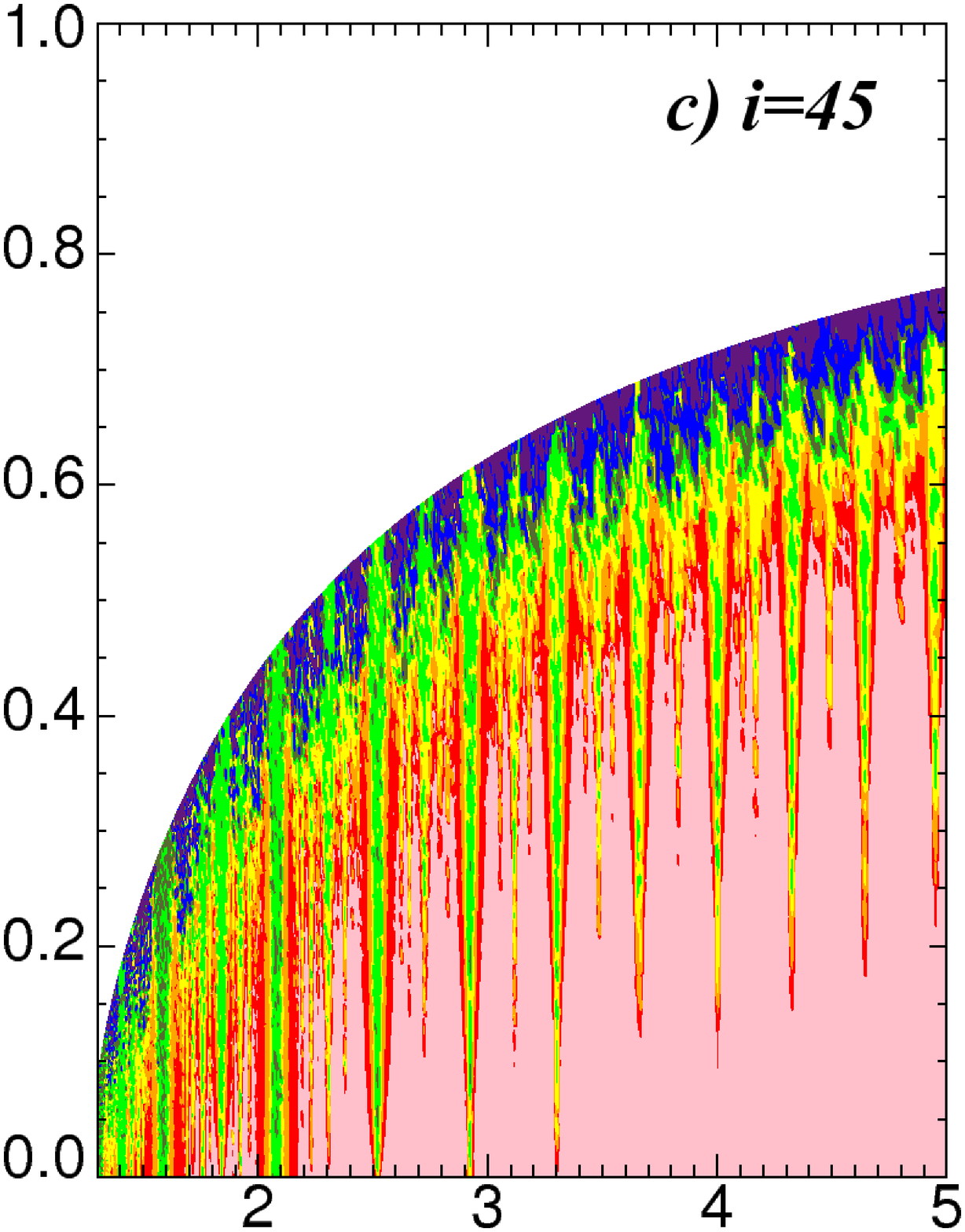}\\
    \includegraphics[angle=0,width=0.33\textwidth]{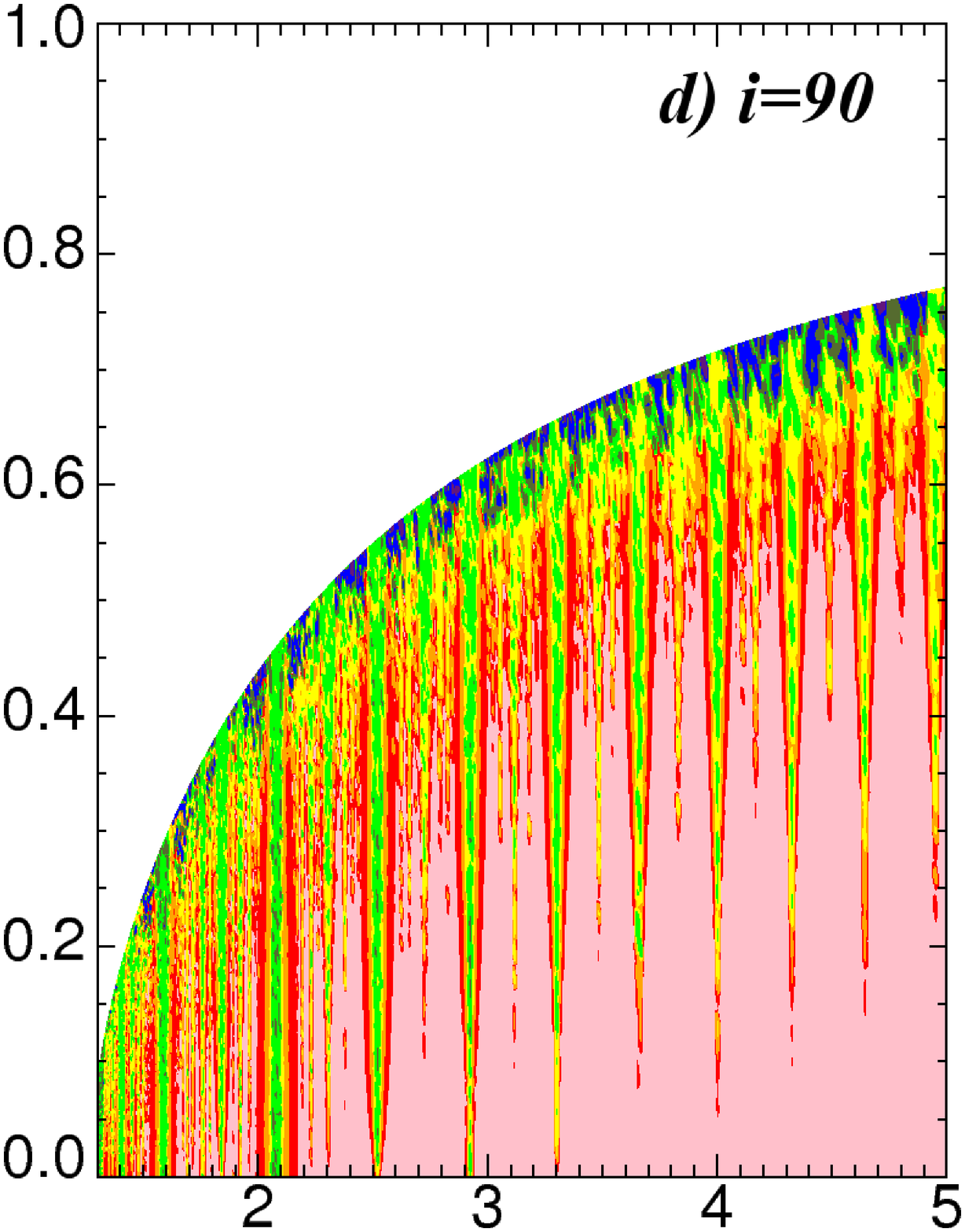}&
    \includegraphics[angle=0,width=0.33\textwidth]{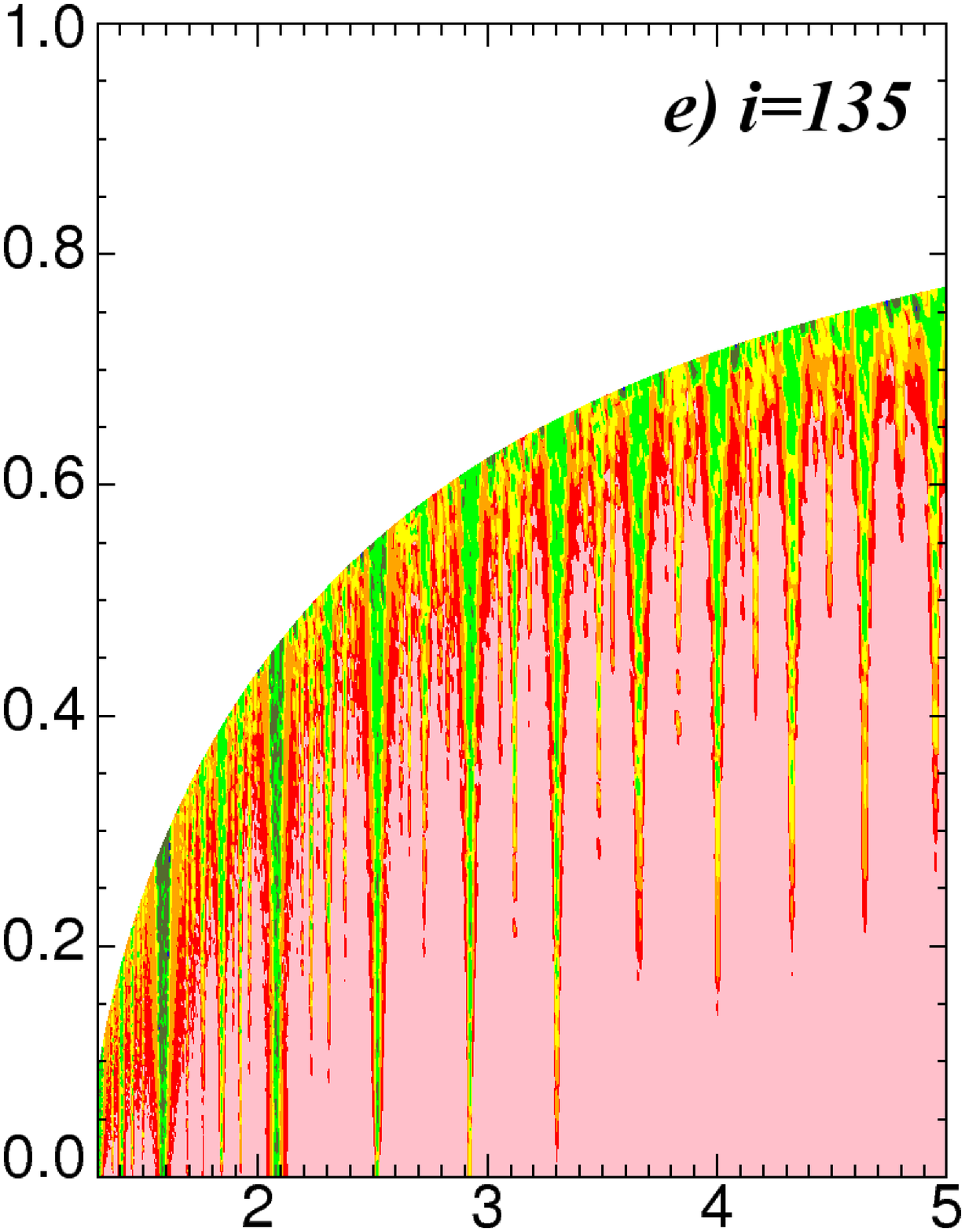}&
    \includegraphics[angle=0,width=0.33\textwidth]{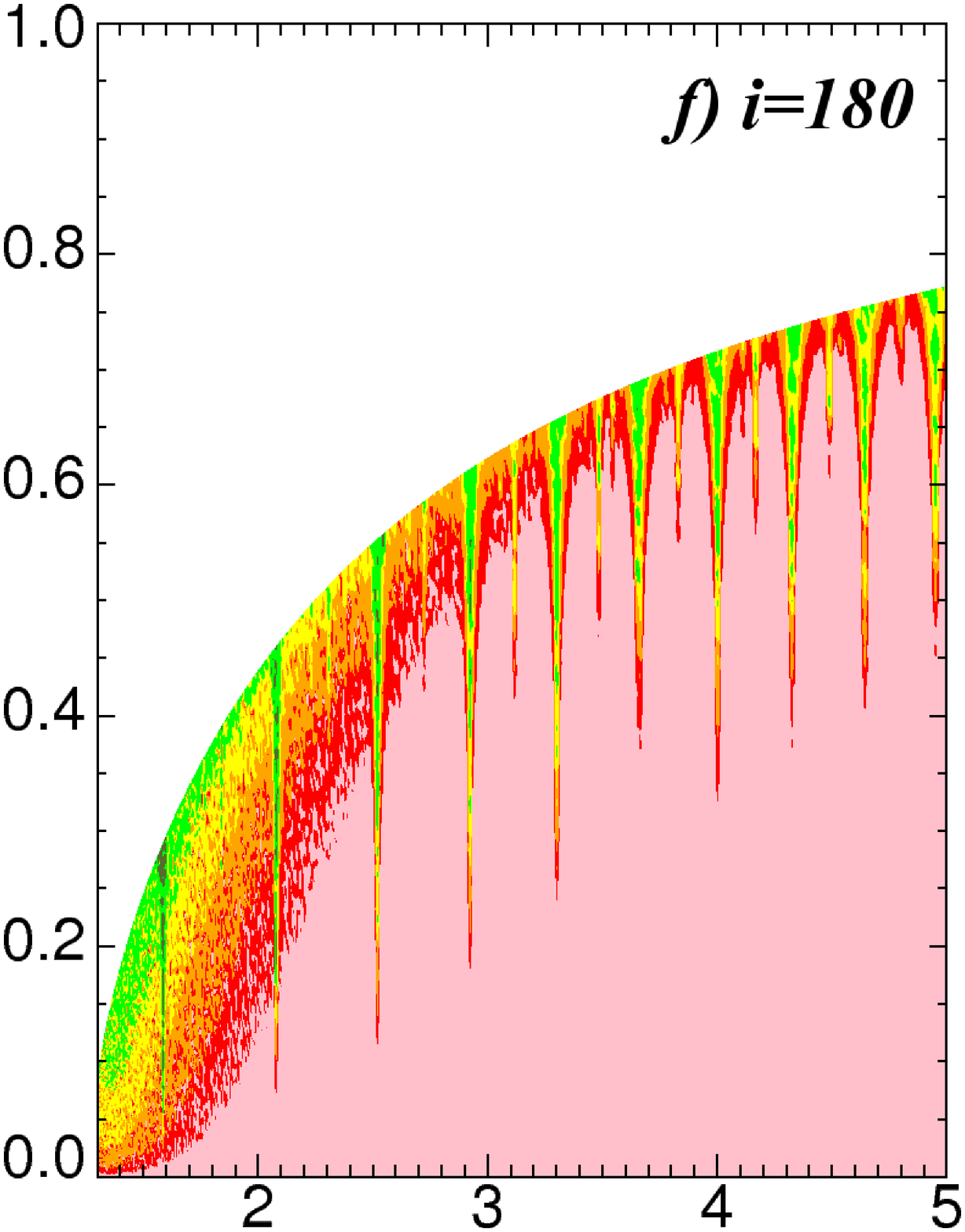}
  \end{tabular}
%
%
  \put(-540,-120){{\begin{sideways}\parbox{10cm}{\centering \bf Eccentricity of the Outer Body ($e_2$)}\end{sideways}}}
  \put(-340,-220){\centering \bf Semi-major Axis Ratio ($a_2 / a_1$).}
  \caption{Median RMS TTV amplitudes in the ($a_2/a_1$,$e_2$) plane for various $i_2$. The contour key is shown at the top-left in plot (a). In each ($a_2/a_1$,$e_2$) plot, we see a strong increase in amplitude (i) at high eccentricity, and (ii) close to MMRs. As the inclination is increased there is an obvious general decrease in TTV amplitude, particularly pronounced for the retrograde orbits.}
  \label{FIG:Fiducial_Inc_1}
\end{figure*}

In \Fig{FIG:Fiducial_Inc_1}a we reproduce the coplanar prograde results of \citet{Agol05}, showing the pronounced ``flames of resonance''  \citep{Veras09c}. The majority of the non-resonant parameter-space has TTV signals of comparatively low amplitude ($< 30$ seconds). It is only when eccentricities become very high, or the planets are in / near MMRs, that the TTV signal amplitudes become large. 

As the relative inclination of the planetary orbits is increased, for low inclinations changes in TTV amplitude are relatively subtle; for the $i=30\,^{\circ}$ \& $i=45\,^{\circ}$ cases (\Fig{FIG:Fiducial_Inc_1}c \& \ref{FIG:Fiducial_Inc_1}d) the RMS TTV signal is reduced in some high-eccentricity locations, corresponding to the reduction previously seen in the close-to-resonance map of \Fig{FIG:Fiducial_Inc_Scan}d. In addition, we can also see that spikes around higher resonances start to drop lower down, i.e. the low eccentricity amplitude \emph{increases} as the inclination decreases, an effect noted in \S \ref{ResultsFixedMaps} for the 5:1 resonance plot of \Fig{FIG:Fiducial_Inc_Scan}c. 

For larger relative inclinations the expected TTV signal amplitude starts to drop across much of the $(\frac{a_2}{a_1},e_2)$ parameter space (see the $i=90\,^{\circ}$ plot in \Fig{FIG:Fiducial_Inc_1}d). The reduction in high amplitude signals at high eccentricities is the most marked decrease, but at any given eccentricity one tends to find a reduction in amplitude (shifting of the contours) compared to the $i=0\,^{\circ}$ plot.

When the systems are pushed into \emph{retrograde} orbits, the difference becomes more pronounced (E.g. $i=135\,^{\circ}$ in \Fig{FIG:Fiducial_Inc_1}e and $i=180$ in \Fig{FIG:Fiducial_Inc_1}f). For these inclinations we find a reduction in amplitude across large swathes of the plot, with significant amplitude remaining only near MMRs. In addition, the TTV amplitude and the width of regions with large TTVs (near MMRs) is markedly reduced.

\subsubsection{External 2:1 MMR}\label{ResultsVaryingExternal}
Since highly inclined systems only have large RMS TTVs in / near MMRs (\Fig{FIG:Fiducial_Inc_1}f), we examine in more detail one such resonant region. The external 2:1 MMR is known to be of importance for exo-planet systems from both theoretical \citep[e.g.][]{Sandor07} and observational \citep[e.g.][]{Tinney06} studies.


\begin{figure*}
  \centering
  \begin{tabular}{ccc}
    \includegraphics[angle=0,width=0.33\textwidth]{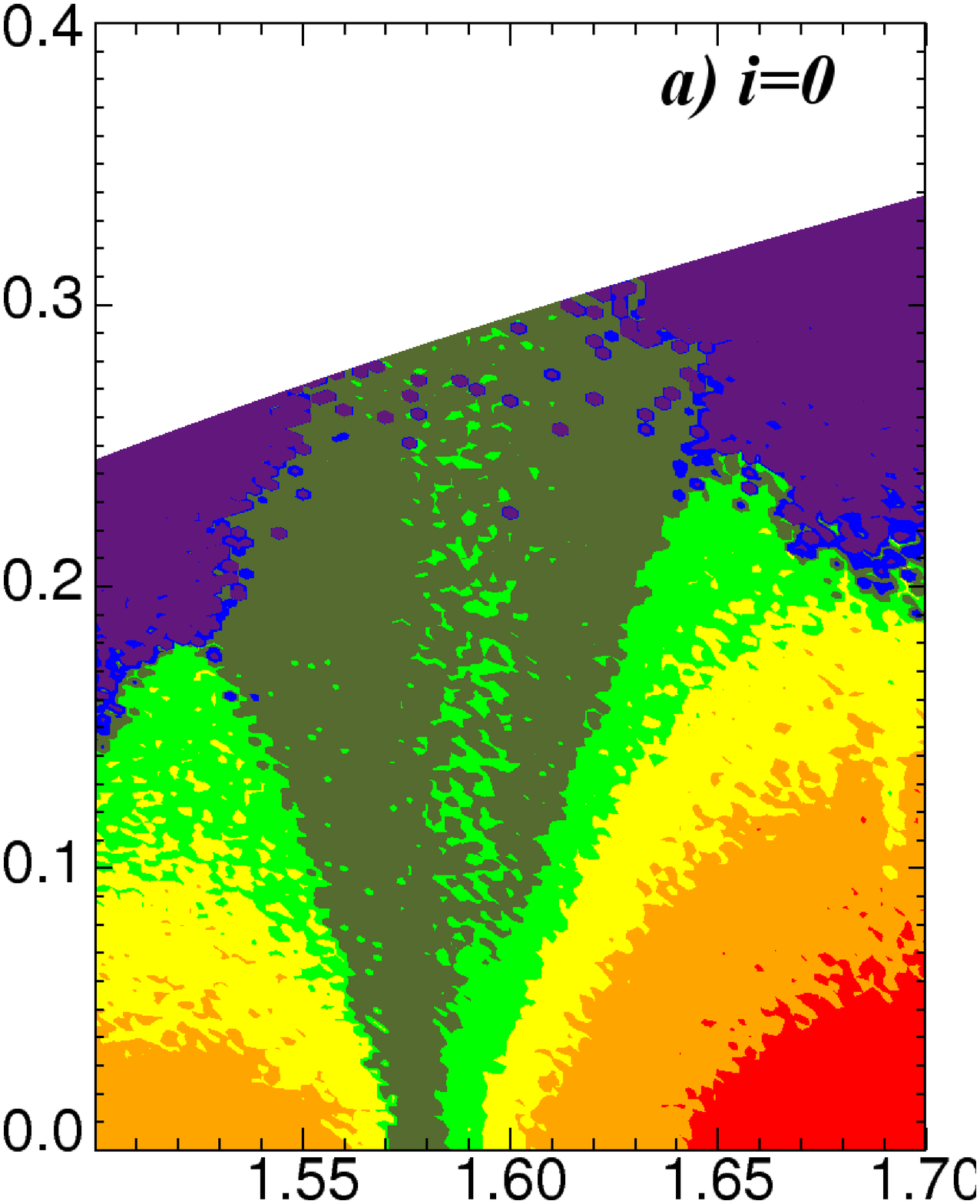}&
    \includegraphics[angle=0,width=0.33\textwidth]{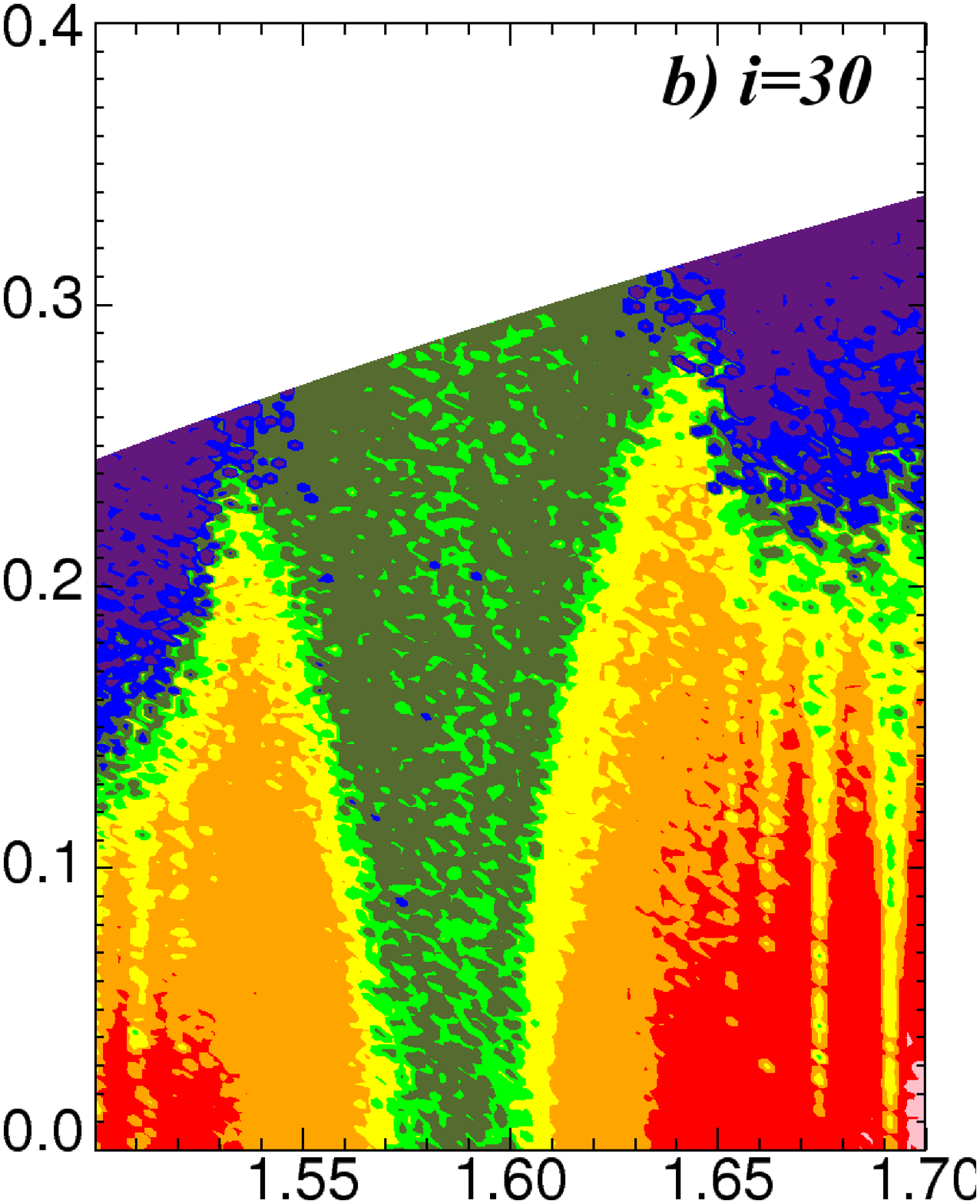}&
    \includegraphics[angle=0,width=0.33\textwidth]{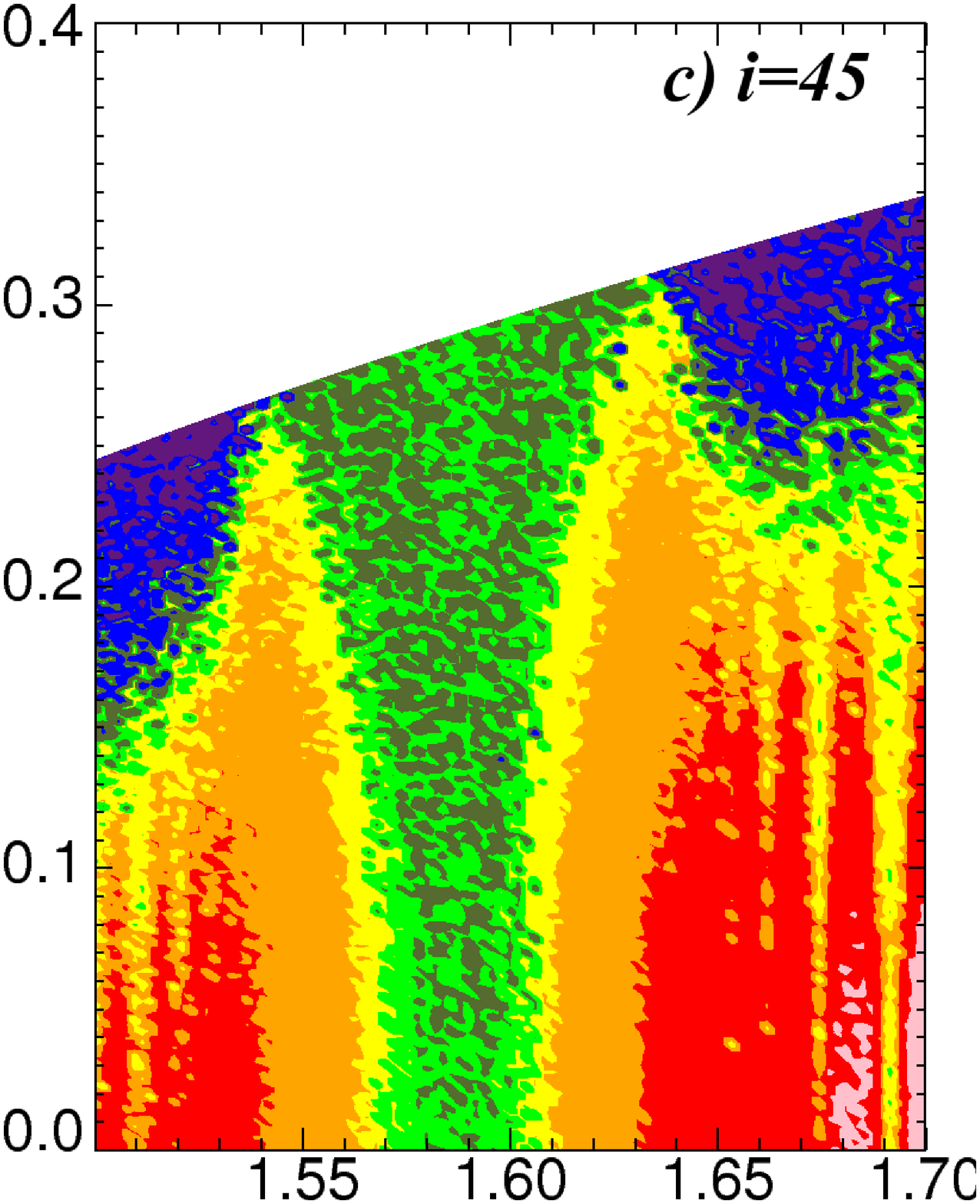}\\
    \includegraphics[angle=0,width=0.33\textwidth]{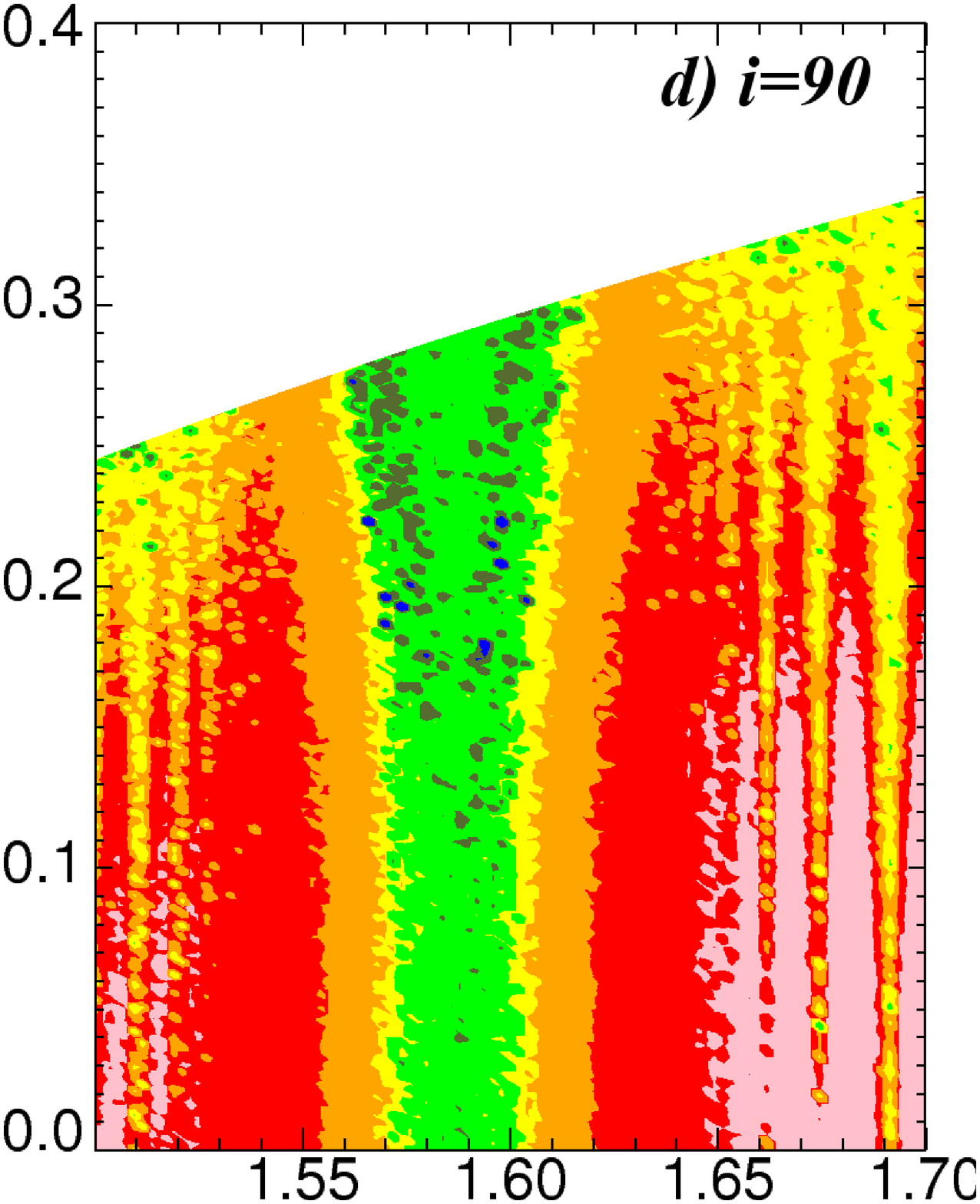}&
    \includegraphics[angle=0,width=0.33\textwidth]{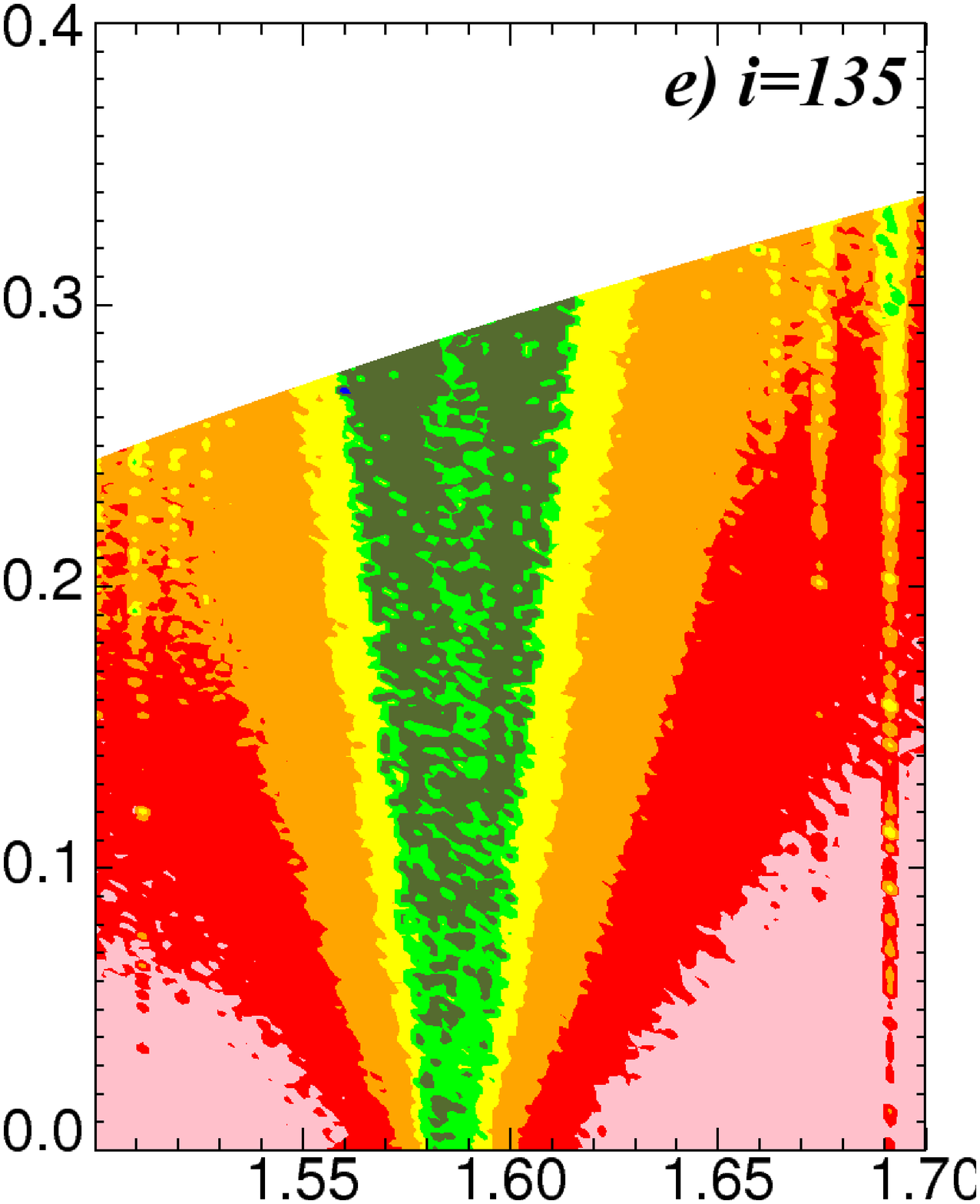}&
    \includegraphics[angle=0,width=0.33\textwidth]{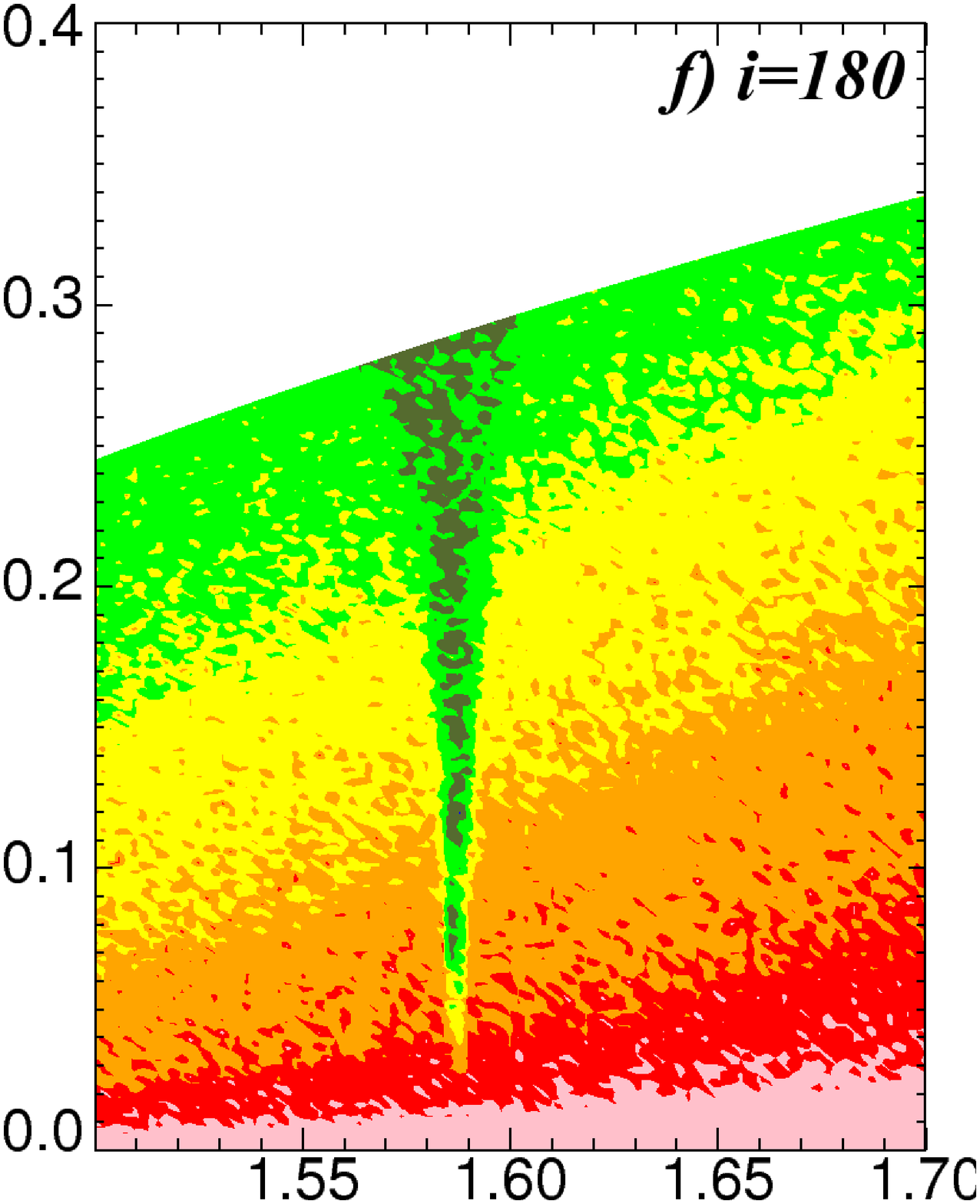}
  \end{tabular}
%
%
  \put(-540,-120){{\begin{sideways}\parbox{10cm}{\centering \bf Eccentricity of the Outer Body ($e_2$)}\end{sideways}}}
  \put(-340,-230){\centering \bf Semi-major Axis Ratio ($a_2 / a_1$).}
  \caption{Close-up of the 2:1 MMR region, demonstrating extreme sensitivity to variations in $a_2/a_1$, $e_2$ \& $i_2$. The contour key is the same as that used in Figs. \ref{FIG:Fiducial_Inc_Scan} \& \ref{FIG:Fiducial_Inc_1}. }
  \label{FIG:Ext_Res}
\end{figure*}

On and adjacent to the 2:1 MMR region, the TTV signal is extremely sensitive to changes in semi-major axis and / or eccentricity. In the co-planar prograde case (\Fig{FIG:Ext_Res}a), a fractional increase of just a few percent in either $a_2$ or $e_2$ causes the TTV amplitude to jump by more than an order of magnitude.

As the inclination increases to $30\,^{\circ}$ and $45\,^{\circ}$, we find that the central region of the MMR maintains an approximately constant TTV amplitude. The regions just outside the resonance feel the strongest effects. E.g. regions near $a_2 / a_1 = 1.54$ show TTV amplitudes reduced by an order of magnitude for many eccentricities (compare \Fig{FIG:Ext_Res}a and \Fig{FIG:Ext_Res}c). This decreases the width of the region of parameter space around the 2:1 MMR resonance with TTV signals $> 30$ seconds. In addition, the general reduction in ``background amplitude'' causes the minor resonances to become more prominent: E.g. $a_2 / a_1 \sim 1.69$.

As the inclination increases to $90\,^{\circ}$ and $135\,^{\circ}$, TTV signals $> 30$ seconds are restricted \emph{purely} to the resonance regions, with regions away from resonance having signals $< 30$ seconds almost everywhere, even at high eccentricity. As the inclination reaches $180\,^{\circ}$, and the planets are now again coplanar but with completely counter-rotating orbits, we find TTV amplitude over 100 seconds only in very narrow regions around the first- and second-order MMRs.

Finally, we note that as we move from \Fig{FIG:Ext_Res}e to \Fig{FIG:Ext_Res}f, the amplitude in regions slightly offset from resonance appears to increase slightly at high eccentricities. However this only occurs at \emph{low} order resonances (small semi-major axis separations - see \Fig{FIG:Fiducial_Inc_1}e and \Fig{FIG:Fiducial_Inc_1}f) as well as \Fig{FIG:Fiducial_Inc_Scan}. More generally, away from such regions (see \Fig{FIG:Fiducial_Inc_1}f), retrograde orbits tend to produce the smallest TTVs.

\subsubsection{Internal Perturbers}\label{ResultsVaryingInternal}
%
\begin{figure*}
  \centering
  \begin{tabular}{cccc}
    \includegraphics[angle=0,width=0.25\textwidth]{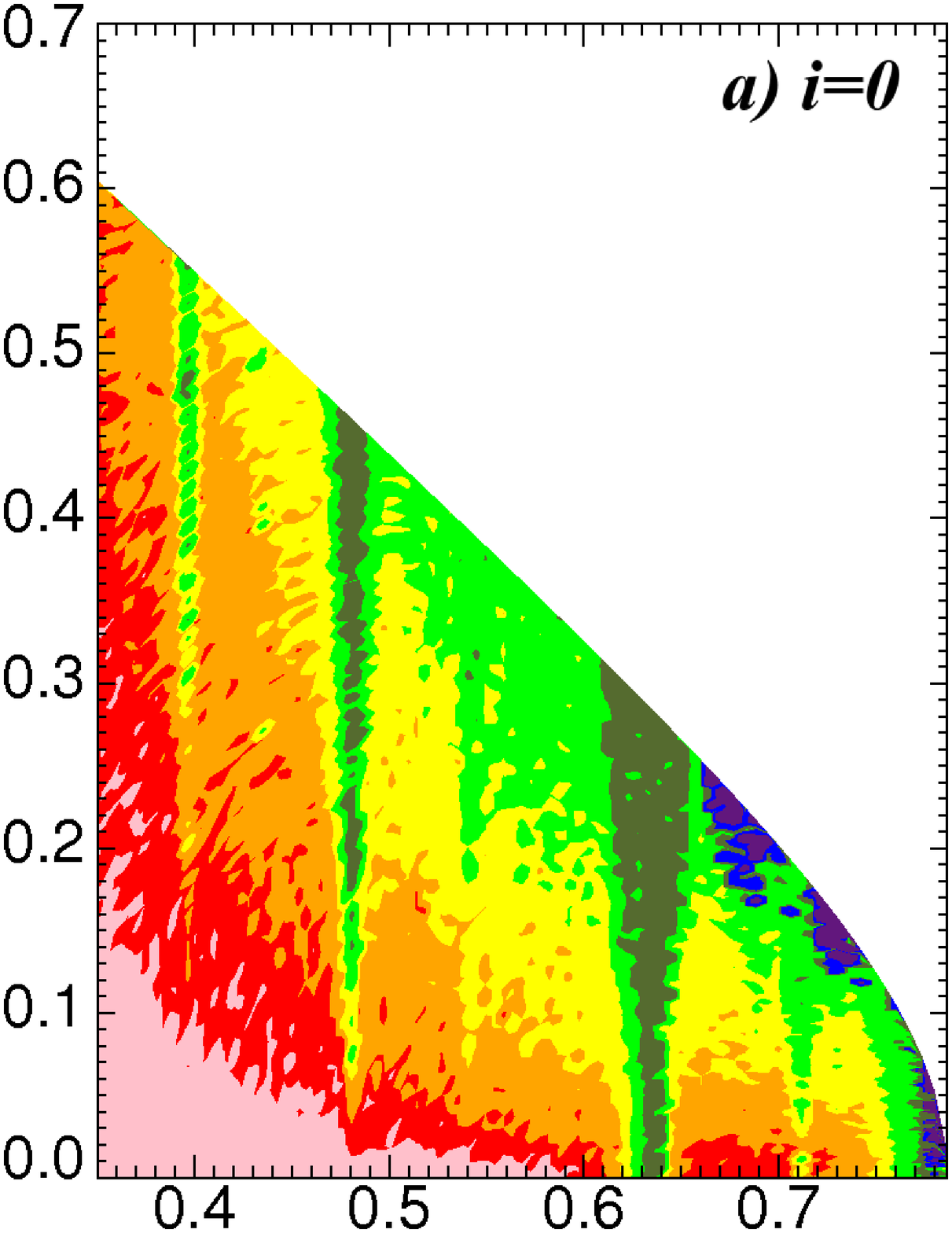}&
    \includegraphics[angle=0,width=0.25\textwidth]{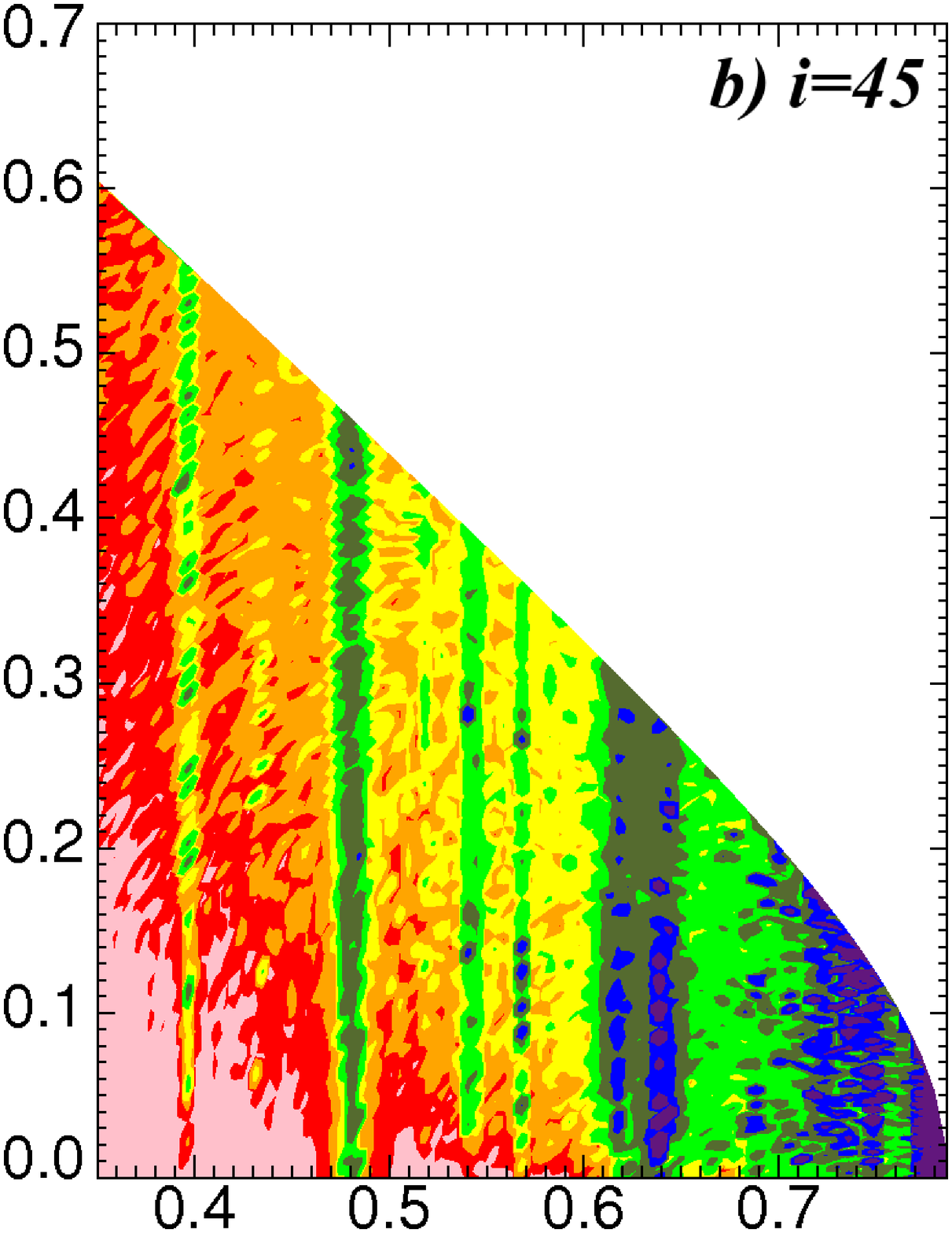}&
    \includegraphics[angle=0,width=0.25\textwidth]{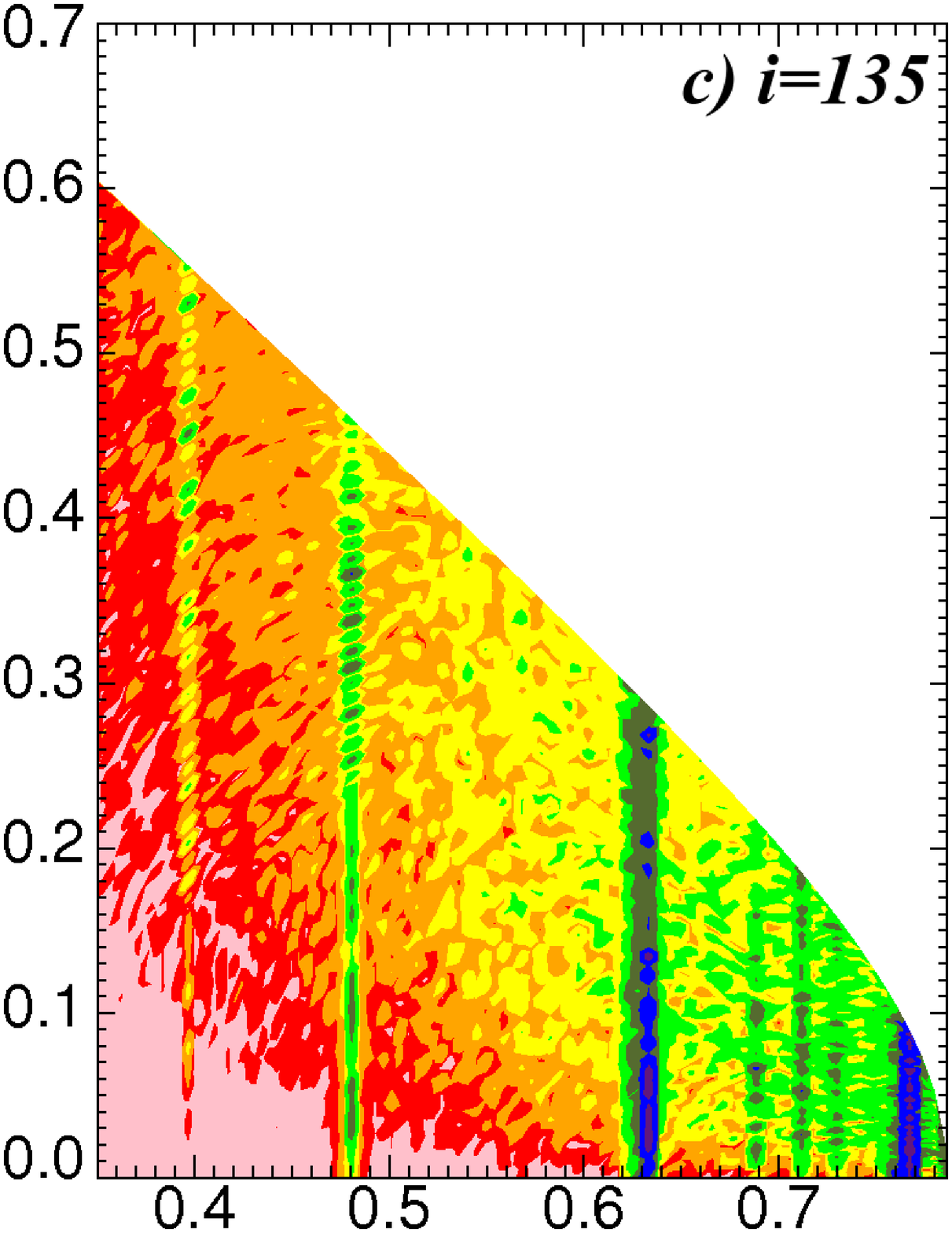}&
    \includegraphics[angle=0,width=0.25\textwidth]{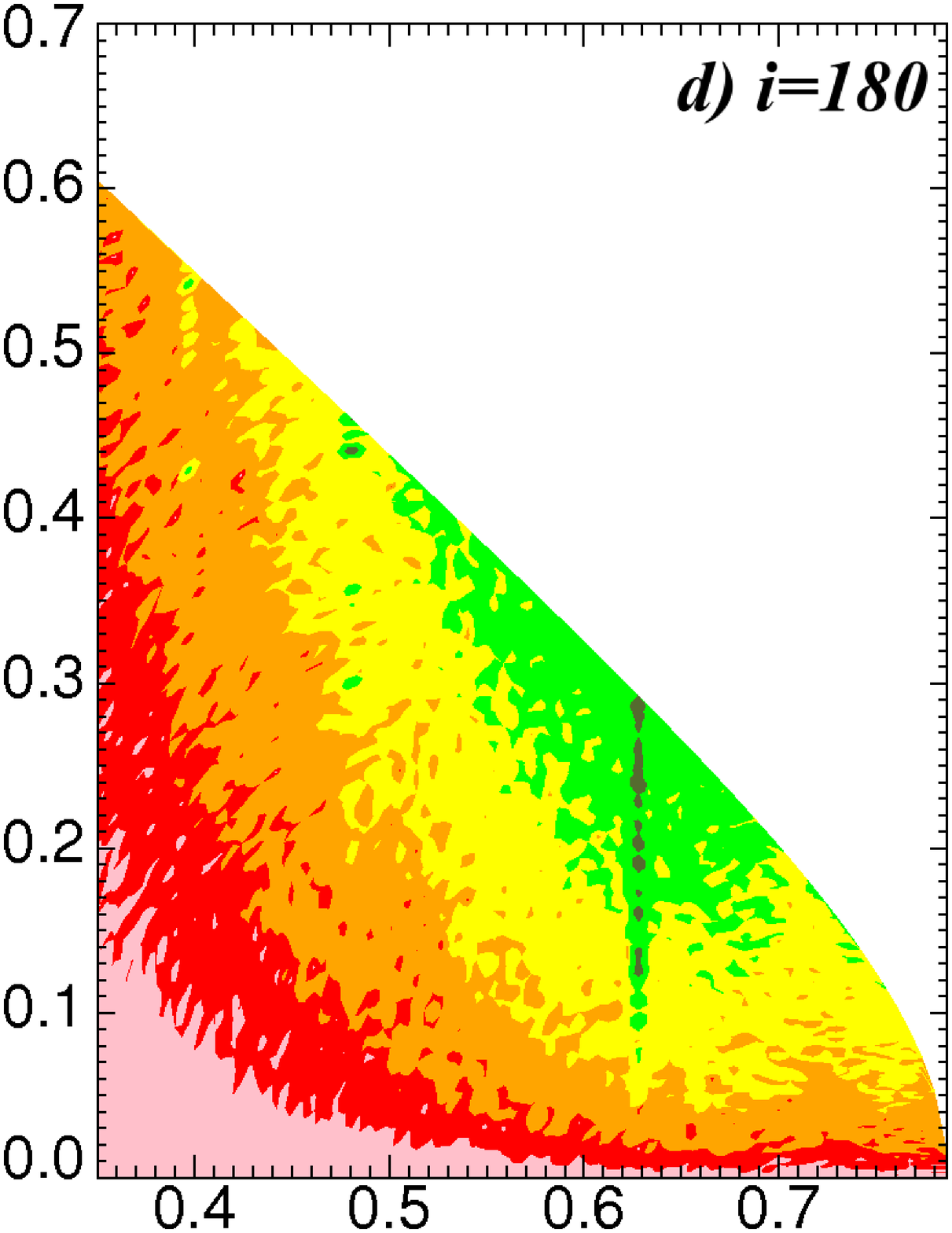}
\\
    \includegraphics[angle=0,width=0.25\textwidth]{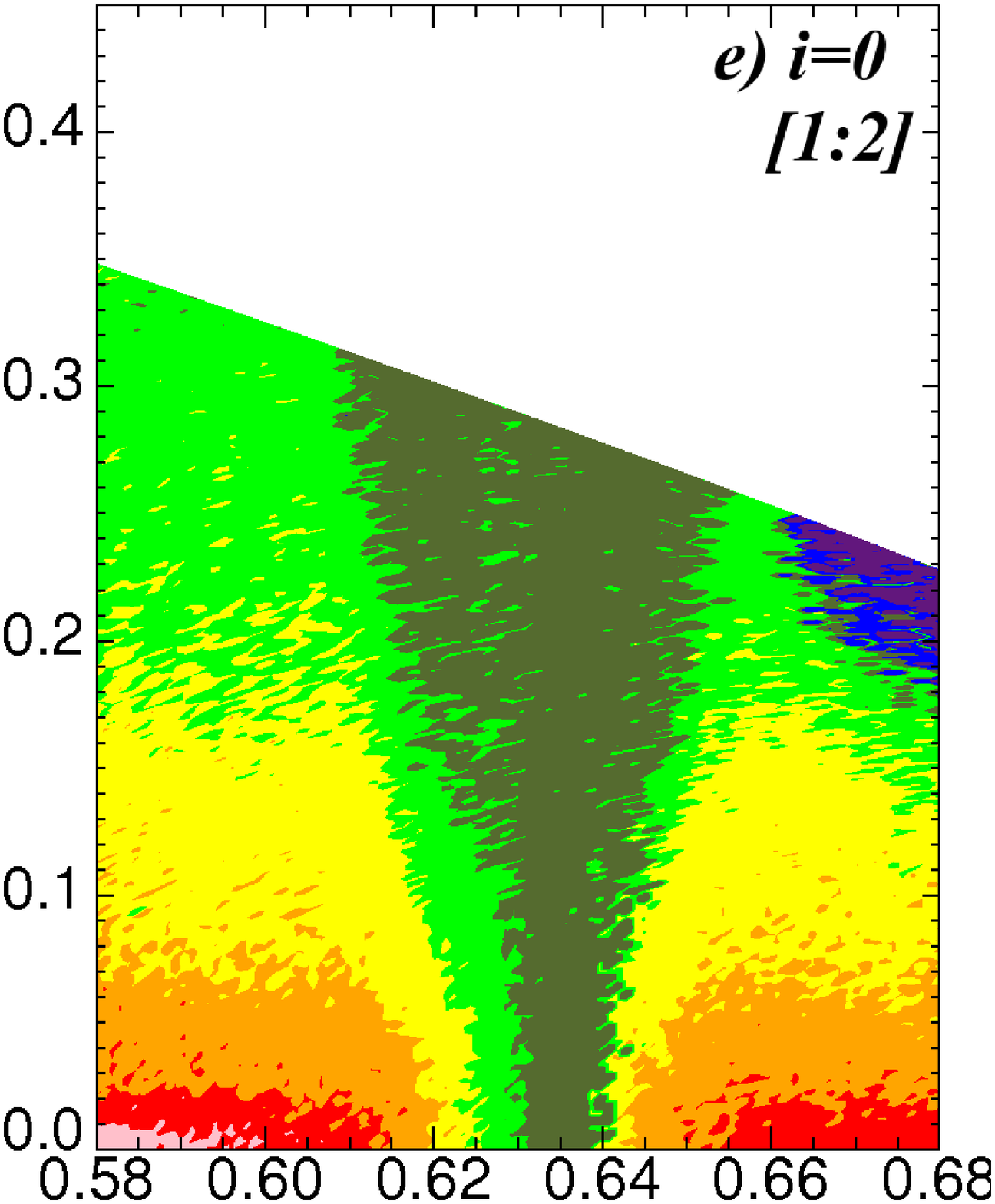}&
    \includegraphics[angle=0,width=0.25\textwidth]{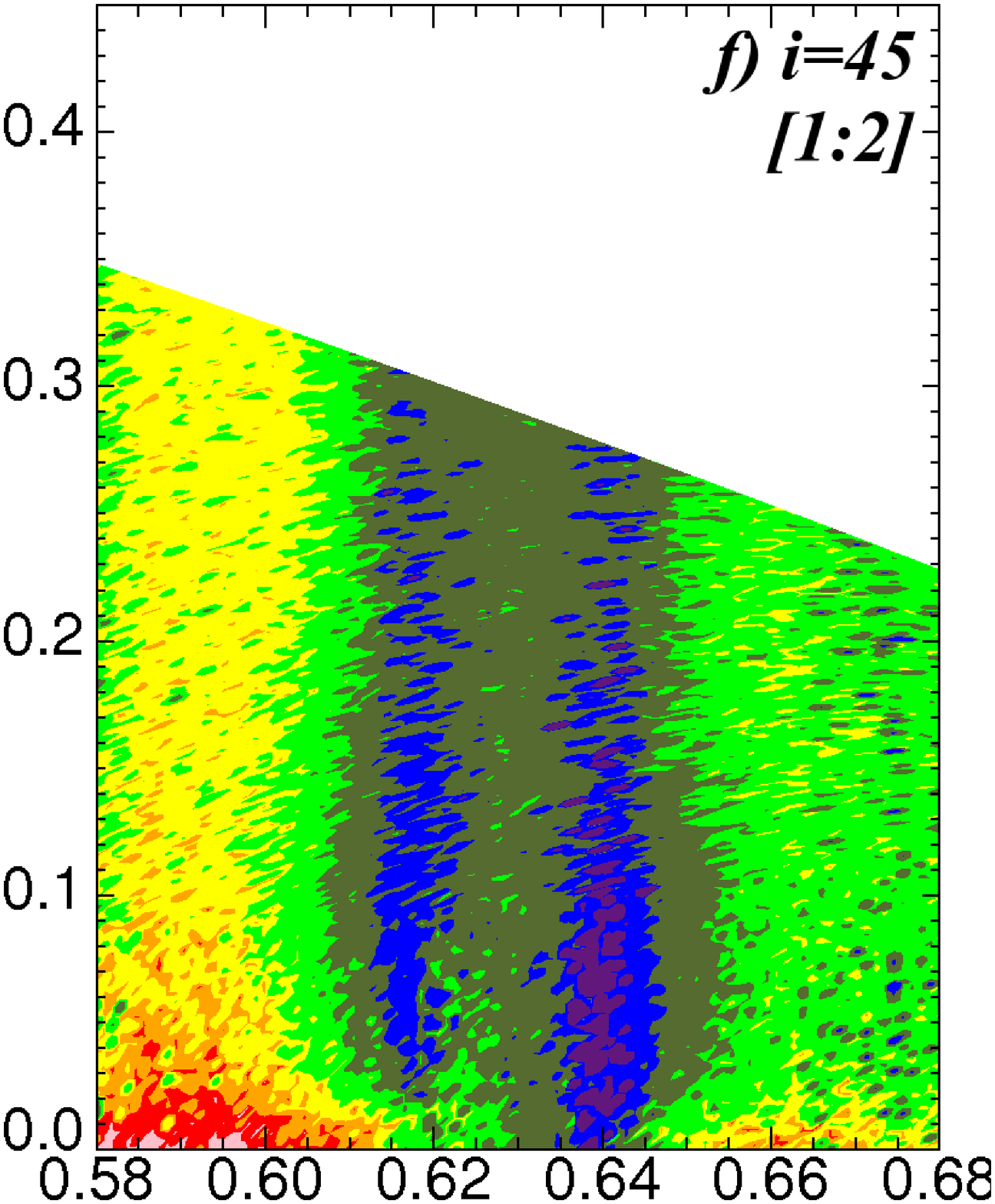}&
    \includegraphics[angle=0,width=0.25\textwidth]{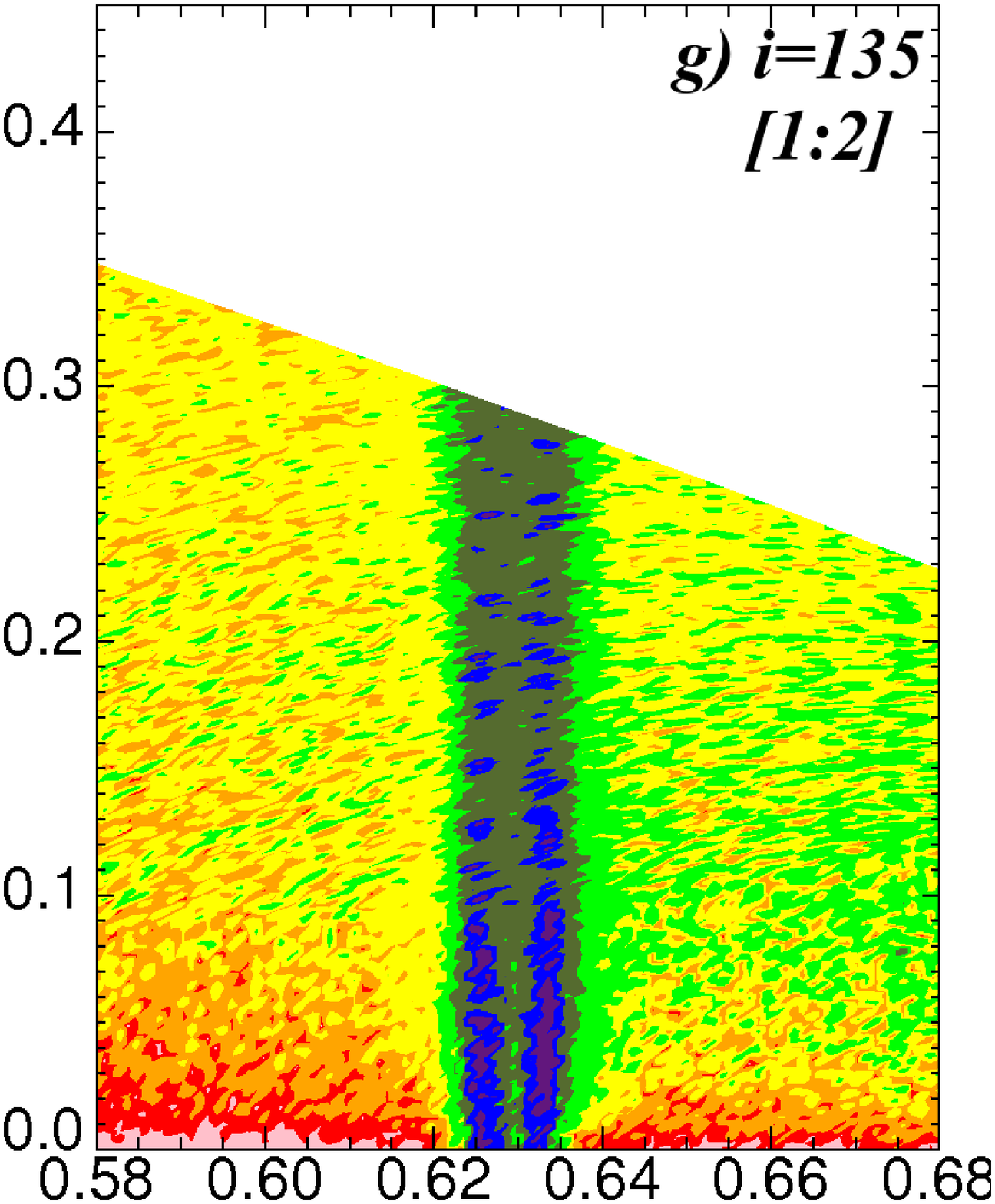}&
    \includegraphics[angle=0,width=0.25\textwidth]{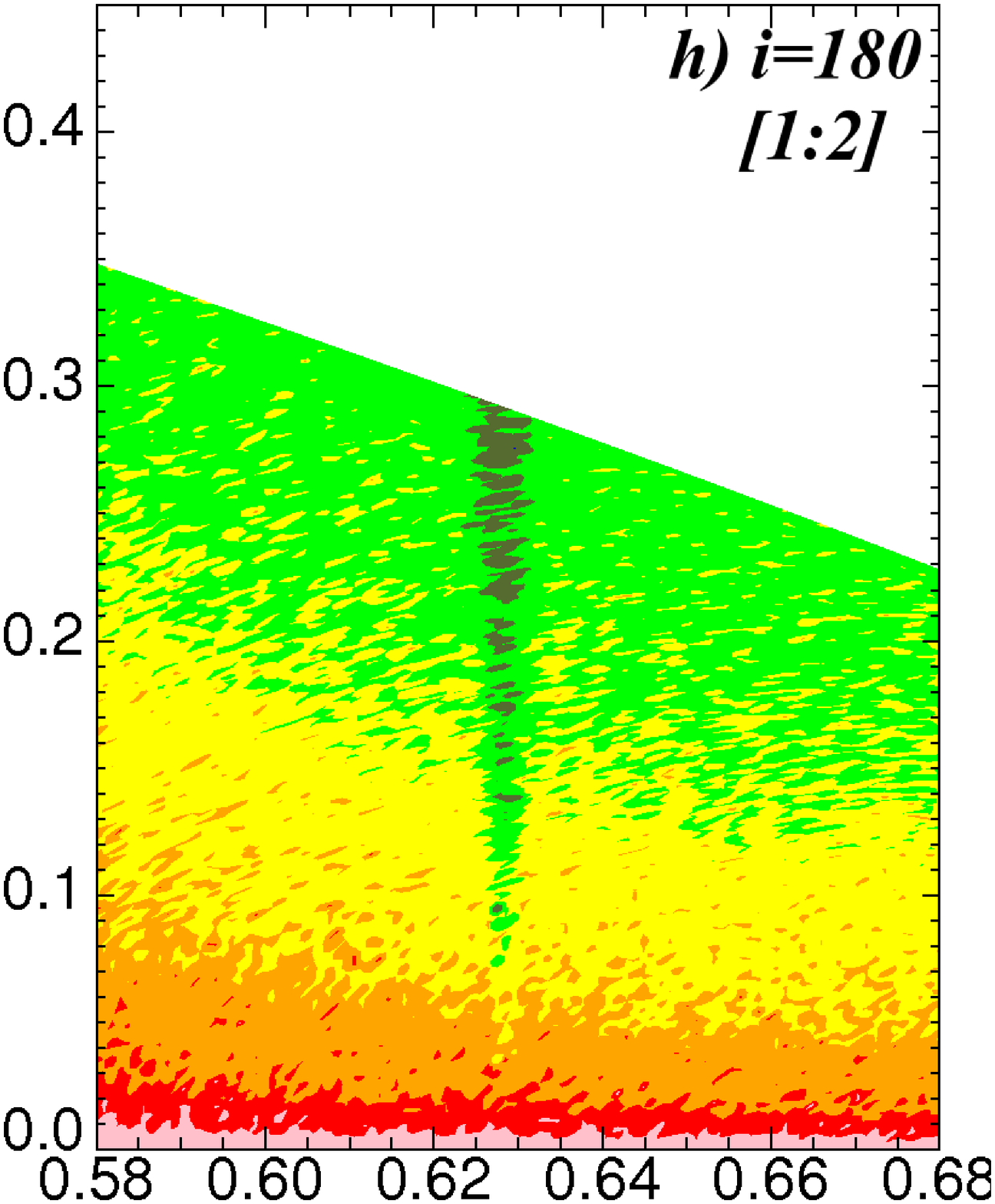}
  \end{tabular}
%
%
  \put(-555,-120){{\begin{sideways}\parbox{10cm}{\centering \bf Eccentricity of the Outer Body ($e_2$)}\end{sideways}}}
  \put(-340,-190){\centering \bf Semi-major Axis Ratio ($a_2 / a_1$).}
  \caption{\emph{Internal} perturber parameter space. Top; a general view. Bottom; a close-up of the 1:2 MMR. The contour key is the same as in \Fig{FIG:Ext_Res}. We find that the behaviour in the internal region differs from the external region: the background stays constant and there is now a \emph{strengthening} of the resonant regions as the inclination increases towards $90\,^{\circ}$.}
  \label{FIG:Int_Res}
\end{figure*}
%
Finally, for completeness we consider an Earth-mass planet on an orbit \emph{interior} to the Jupiter-mass planet, keeping the more massive planet at $a_1 = 0.05 AU$ and $e_1 = 0$, and then monitor the TTVs of the massive planet due to the interior perturber.

From \Fig{FIG:Int_Res} it seems that the background (non-resonant) regions now retain an approximately constant TTV signal irrespective of inclination, while there is a tendency for the TTV signal to become \emph{stronger} around the MMR regions as the inclination is increased towards $90\,^{\circ}$, but then drop away to almost nothing as the inclination further increases towards $i=180\,^{\circ}$.


\section{Conclusions}
We have investigated the TTV signals for systems of highly inclined and retrograde planets. We find that:
\begin{itemize}
\item In the vicinity of exterior MMRs the inclination dependence is complex: low order resonances maintain a high TTV amplitude for all regions $i<170\,^{\circ}$, declining in amplitude only for low eccentricity cases close to $i=180\,^{\circ}$, where-as higher order resonances display an increase in TTV amplitude as inclinations rise from 0 to 45 degrees. Moreover, the regions immediately adjacent to MMRs show extreme sensitivity to changes in perturber $a$, $e$ and $i$.
\item Exterior perturbers away from resonances tend to show a slow decrease in TTV amplitude with increasing inclination, although regions adjacent to resonances can show remarkably complex behaviour.
\item Perturbing planets on interior orbits display a slightly different behaviour: Away from resonance the amplitude remains approximately constant with inclination, but around MMRs the perturbations become \emph{stronger} as the inclination increases towards $90\,^{\circ}$ before decreasing again beyond $90\,^{\circ}$.
\end{itemize}

We note that:
\begin{itemize}
\item Absence of evidence is not evidence of absence: Planets in retrograde orbits should be expected to produce markedly reduced TTV signals as compared to the standard prograde case. For an Earth-mass perturber in an anti-aligned orbit, almost the entirety of the sample parameter space would result in a very small or undetectable TTV signal (unless the planet happened to be fortuitously located precisely on a strong MMR).
\item Retrograde orbits may be a natural way to explain transiting systems in which little or no TTV signal is observed, but in which the radial velocity observations point towards the existence of an additional planetary companions (E.g. GJ 436, HAT-P-13, etc).
\item In addition to the TTV considerations in this work, it is important to acknowledge that inclined orbits in multi-planet systems will precess, leading to variations in transit duration. More work is required to try and understand whether a combination of TTV signals with transit \emph{duration} variation signals could remove some of the degeneracies inherent in this problem.
\item Many additional dependencies (E.g. sampling period, perturber mass, relative mean anomaly, etc, etc) can significantly alter the expected TTV signal for a particular system. We defer the provision of a detailed investigation of such matters to a companion paper, \emph{Veras et al. 2010, in prep.}
\item Future work could also investigate in more detail the regions \emph{above} the Hill stability curve, seeking to identify the (likely large) TTVs for any stable systems which exist in that region.
\end{itemize}


%
\acknowledgments
This material is based upon work supported by the National Science Foundation under Grant No. 0707203 and is also based on work supported by the National Aeronautics and Space Administration under grant NNX08AR04G issued through the Kepler Participating Scientist Program. We are very grateful to Eric Agol and our anonymous referee for providing detailed comments which helped to significantly clarify sections of this letter. 

\end{document}